\documentclass[sigconf,authorversion,nonacm]{acmart}
\usepackage[utf8]{inputenc}
\usepackage{xspace}
\usepackage{multirow}
\usepackage{cleveref}
\usepackage{subcaption}

\usepackage[inline]{enumitem}
\usepackage{bbm}
\usepackage{amsmath}
\usepackage{amsfonts}
\usepackage[autolanguage]{numprint}
\usepackage{booktabs}
\usepackage{diagbox}

\newcommand{\citeauthorwitcite}[1]{\citeauthor{#1}~\cite{#1}}

\newcommand{\eg}{e.g.,\xspace}
\newcommand{\ie}{i.e.,\xspace}

\newcommand{\furl}[1]{\footnote{\url{#1}}}

\newcommand{\bertforrec}{BERT4Rec\xspace}
\newcommand{\narm}{NARM\xspace}
\newcommand{\gru}{GRU\xspace}
\newcommand{\sasrec}{SASRec\xspace}

\newcommand{\hr}[1]{HR@#1}
\newcommand{\hrat}[1]{HR@#1}
\newcommand{\recallat}[1]{Recall@#1}
\newcommand{\ndcg}[1]{NDCG@#1}
\newcommand{\ndcgat}[1]{NDCG@#1}

\newcommand{\movielenssystem}{Movielens\xspace}
\newcommand{\movielens}{ML-1m\xspace}
\newcommand{\movielenslarge}{ML-20m\xspace}
\newcommand{\ambeauty}{Amazon Beauty\xspace}
\newcommand{\amgame}{Amazon Games\xspace}
\newcommand{\steam}{Steam\xspace}

\newcommand{\exact}{full\xspace}
\newcommand{\Exact}{Full\xspace}
\newcommand{\kendall}{Kendall's Tau\xspace}
\newcommand{\kendallc}{Kendall's Tau correlation coefficient\xspace}
\newcommand{\targetsize}{sample size\xspace}
\newcommand{\targetsizes}{sample sizes\xspace}
\newcommand{\its}{target set\xspace}

\copyrightyear{2021} 
\acmYear{2021} 
\setcopyright{acmlicensed}\acmConference[RecSys '21]{Fifteenth ACM Conference on Recommender Systems}{September 27-October 1, 2021}{Amsterdam, Netherlands}
\acmBooktitle{Fifteenth ACM Conference on Recommender Systems (RecSys '21), September 27-October 1, 2021, Amsterdam, Netherlands}
\acmPrice{15.00}
\acmDOI{10.1145/3460231.3475943}
\acmISBN{978-1-4503-8458-2/21/09}

\begin{document}

\title[Case Study on Sampling Strategies for Evaluating Sequential Recommendation]{A Case Study on Sampling Strategies for Evaluating Neural Sequential Item Recommendation Models}

\author{Alexander Dallmann}
\email{dallmann@informatik.uni-wuerzburg.de}
\authornote{Both authors contributed equally to this research.}
\author{Daniel Zoller}
\authornotemark[1]
\email{zoller@informatik.uni-wuerzburg.de}
\author{Andreas Hotho}
\email{hotho@informatik.uni-wuerzburg.de}
\affiliation{%
  \institution{University of Würzburg}
  \department{Data Science Chair}
  \streetaddress{Am Hubland}
  \city{Würzburg}
%  \state{Bavaria}
  \country{Germany}
  \postcode{97074}
}

\begin{abstract}
At the present time, sequential item recommendation models are compared by calculating metrics on a small item subset (\textit{\its}) to speed up computation.
The \its contains the relevant item and a set of negative items that are sampled from the full item set.
Two well-known strategies to sample negative items are \textit{uniform random sampling} and \textit{sampling by popularity} to better approximate the item frequency distribution in the dataset.
Most recently published papers on sequential item recommendation rely on \textit{sampling by popularity} to compare the evaluated models.
However, recent work has already shown that an evaluation with \textit{uniform random sampling} may not be consistent with the \exact ranking, that is, the model ranking obtained by evaluating a metric using the full item set as \its, which raises the question whether the ranking obtained by sampling by popularity is equal to the \exact ranking.
In this work, we re-evaluate current state-of-the-art sequential recommender models from the point of view, whether these sampling strategies have an impact on the final ranking of the models.
We therefore train four recently proposed sequential recommendation models on five widely known datasets.
For each dataset and model, we employ three evaluation strategies.
First, we compute the \exact model ranking. 
Then we evaluate all models on a \its sampled by the two different sampling strategies, \textit{uniform random sampling} and \textit{sampling by popularity} with the commonly used \its size of $100$, compute the model ranking for each strategy and compare them with each other.
Additionally, we vary the size of the sampled \its.
Overall, we find that both sampling strategies can produce inconsistent rankings compared with the full ranking of the models.
Furthermore, both sampling by popularity and uniform random sampling do not consistently produce the same ranking when compared over different sample sizes.
Our results suggest that like uniform random sampling, rankings obtained by sampling by popularity do not equal the \exact ranking of recommender models and therefore both should be avoided in favor of the \exact ranking when establishing state-of-the-art.
\end{abstract}

\keywords{Sequential Item Recommendation, Evaluation, Metrics, Sampled Metrics}

\begin{CCSXML}
<ccs2012>
<concept>
<concept_id>10002951.10003317.10003347.10003350</concept_id>
<concept_desc>Information systems~Recommender systems</concept_desc>
<concept_significance>500</concept_significance>
</concept>
<concept>
<concept_id>10002951.10003317.10003359</concept_id>
<concept_desc>Information systems~Evaluation of retrieval results</concept_desc>
<concept_significance>500</concept_significance>
</concept>
</ccs2012>
\end{CCSXML}

\ccsdesc[500]{Information systems~Recommender systems}
\ccsdesc[500]{Information systems~Evaluation of retrieval results}

\maketitle

\section{Introduction}
\label{sec:introduction}

A crucial part of developing recommender systems is the evaluation of the model candidates during the process.
Online evaluation is still the best choice for evaluating recommender models~\cite{liang2018variational}, but it is not applicable during the development of a new model (\eg for finding the best hyper-parameter settings).
So, offline evaluation remains the best option for the recommender community to evaluate new models during development.
Because recommender systems must cover increasingly larger areas of application (in terms of number of items) in recent years, \citeauthorwitcite{koren2008factorization} introduced an evaluation procedure to speed up the process of metric calculation during evaluation on these larger item sets.
Instead of computing the metrics on the full item set to get a ranking (called \textit{full ranking}) for the models, the metrics are computed on a target set, that is a small subset of items, containing all relevant items and a defined number of negative (non-relevant) items that are sampled \textit{uniform} from the full item set.
Many scientific work has adapted the method by mainly changing the number of negative samples in the target set (\eg~\cite{elkahky2015multiview, he2017neural, kang2018selfattentive}).
Others have changed the process by sampling the item set based on the \textit{popularity} of the items~\cite{sun2019bert4rec} to make the sampling more representative and reliable.
It became common practice to use the sampling not only while training or developing the model, but also for reporting the performance of the recommender models~\cite{koren2008factorization, kang2018selfattentive, sun2019bert4rec}.

Recently, in~\cite{Krichene20} the authors formally showed that the expected values of most utilized metrics for recommender evaluation based on \textit{uniform sampling} depend on the rank of the relevant item assigned by the model under evaluation.
Therefore, the ranking on the sampled target set can differ from the one on the full item set.
Attempts to correct the metrics were made by \citeauthorwitcite{Krichene20} and also by \citeauthorwitcite{Dong20} in respect to the Hit Rate metric.
Since current state-of-the-art models for sequential recommendation, that use neural networks to extract the user's preference, have also been evaluated using these sampling strategies, we aim to re-evaluate the results of the past years with respect to how their performance ranking changes under the different sampling options.
Thereby, we want to validate and extend the findings of~\cite{Krichene20} to these types of recommender models.
In particular, we focus on \textit{sampling by popularity} for target set generation, because its effects on comparative ranking with other models has not yet been studied.
To that end, we train four neural sequential recommendation models, namely, \gru~\cite{tan2016improved}, a Recurrent Neural Network (RNN), \narm~\cite{li2017neural}, a RNN with attention, \sasrec~\cite{kang2018selfattentive} and \bertforrec~\cite{sun2019bert4rec}, both transformer-based models on five commonly used datasets \steam, Amazon Beauty and Games and \movielenssystem \movielens and \movielenslarge.
Using the trained models, we provide a comparative evaluation of model performance rankings obtained using three different evaluation strategies \begin{enumerate*}
    \item full item set,
    \item sampling by popularity and
    \item uniform random sampling
\end{enumerate*}.
Furthermore, we also investigate changes in rank for different sizes of the sampled target set.
In summary, our main contributions are:
\begin{enumerate}
    \item We are the first, that analyze the effects of sampling the target set by popularity on the consistency with the \exact ranking and the ranking obtained by sampling the target set uniform random.

    \item We re-evaluate four current state-of-the-art models for sequential recommendation on five commonly used datasets to confirm the previously reported ranking achieved with sampling the target set by popularity and compare this ranking with the \exact ranking.
    
    \item We test, whether sampling the target set uniform random yields an inconsistent model ranking with the \exact ranking as reported by~\citeauthorwitcite{Krichene20} on different models and datasets compared to them. 

    \item We examine experimentally the consistency between the \exact ranking and the ranking obtained by sampling on different target set sizes.
\end{enumerate}

In our experiments we can reproduce the results of the previously reported rankings of sequential recommender models using sampling the target set by popularity.
However, our results also show that both rankings obtained by sampling are inconsistent with the \exact ranking on all datasets we tested.
We thereby affirm the reported results of~\citeauthorwitcite{Krichene20} for more datasets and neural sequential recommender models.
When varying the target set size on the datasets, we find that the sampled rankings are not consistent with the \exact ranking.
Overall, our results suggest that the \exact ranking should be used when comparing model performance.

The remainder of this paper is structured as follows:
First, we layout the general setting in \Cref{sec:setting}.
Then we discuss related work in \Cref{sec:relatedwork}
After describing the experimental setup in \Cref{sec:setup}, we present the obtained results of our experiments in \Cref{sec:results}.
Before we conclude the paper in \Cref{sec:conclusion}, we discuss our findings in \Cref{sec:discussion}.
\section{Settings}
\label{sec:setting}
In this section we first specify the sequential item recommendation task. 
Then we formally introduce the evaluation setup and metrics we used to score the recommender models. 
Furthermore, we define the strategies for sampling a \its, that we investigate in this paper.

\newcommand{\sequencesymbol}{s}
\newcommand{\sequenceset}{S}
\newcommand{\itemset}{I}
\newcommand{\hiddensize}{d}
\newcommand{\embeddingsize}{e}
\newcommand{\ntime}[1]{#1_{\sequencelength}}
\newcommand{\ntimeprev}[1]{#1_{t-1}}
\newcommand{\sequence}{\boldsymbol{\sequencesymbol}}
\newcommand{\sequencelength}{l_{\sequence}}
\newcommand{\specifictime}[2]{#1_#2}
\newcommand{\looseqlength}{l_{\sequencesymbol}}

\subsection{Sequential Item Recommendation Task}

The goal of a sequential item recommendation model is to learn a user’s preferences based on her history of item interactions (\eg rating a movie) and to recommend new relevant items based on the accumulated information.
Formally, let $\itemset = \{i_1, i_2, \ldots, i_{|I|}\}$ be the set of items and $U = \{ u_1, u_2, \ldots, u_{|U|} \}$ the set of users.
We construct sequences $s^u = (s^u_1, s^u_2, \ldots ,s^u_{l_{s^u}}) \in S$ with length $l_{s^u}$ and $s^u_j \in I$ for each user $u$.%
\footnote{Since we only leverage the user information to build the interaction sequence, we drop the superscript $u$ in the following for better readability.}
A sequential recommendation model $M: \sequenceset \to R_n$, with $R_n$ being the set of all permutations of the position list $(1, 2, \ldots, |I|)$, is now tasked with ranking the next item first in the position list, given a sequence of past interactions. 
For later, we define the function $\text{head}(s, t) = (\sequencesymbol_1, \sequencesymbol_2, \ldots, \sequencesymbol_t)$, that returns the first $t$ steps of a sequence $s$ and the function $\text{set}(s)$, that converts a sequence $s$ to a set containing all items of sequence~$s$.

\subsection{Dataset Split Strategy}
We use the leave-one-out evaluation strategy as is common in most related work (\eg \cite{sun2019bert4rec, kang2018selfattentive}). 
Therefore, for training we extract for each sequence in $S$ the subsequence containing all items except the two last items: $S_\text{train} = \{\text{head}(s, l_s - 2) \mid s \in S \}$.
The model is validated on the penultimate item $\sequencesymbol_{\looseqlength - 1}$ and later tested on the last item $\sequencesymbol_{\looseqlength}$ for each sequence in the dataset using the metrics defined in the next section.

\newcommand{\sequenceformodel}{\hat{s}}
\subsection{Evaluation Metrics}
We use two common evaluation metrics for evaluating the models.
Given a sequence $s$ and relevant target items $i_r = s_{l_s - 1}$ and $i_r = s_{l_s}$ during validation and testing respectively, the hit rate at position k (\hrat{k})~\cite{shani2011evaluating} measures if the relevant item is in the head/top $k$ of the returned ranking of model $M$ given the sequence $\sequenceformodel = \text{head}(s, l_s - 2)$ for validation and $\sequenceformodel = \text{head}(s, l_s - 1)$ for testing:
\begin{equation}
    \text{\hrat{k}} = \left | \{m_i \in \text{head}(M(\sequenceformodel), k) \mid m_i = r \} \right|.\footnote{Because the recommendation model $M$ returns the item identifiers, we compare these identifiers with the identifier $r$ of the relevant item.}
\end{equation}
Note, that the \hrat{k} is equal to the \recallat{k} in our setting with only one relevant item.
The second metric we use for our experiments is the Normalized Discounted Cumulative Gain (NDCG)~\cite{shani2011evaluating}, that can be formalized in our setting as:
\begin{equation}
    \text{\ndcgat{k}} = \sum_{i=1}^k \delta(M(\sequenceformodel)_i, r) \frac{1}{\log(i + 1)},
\end{equation}
where $M(\sequenceformodel)_i$ is the i-th entry of the ranking and $\delta(a, b) = 1$ if $a = b$ otherwise $0$.

\subsection{Ranking of Recommendation Models}

To rank a set of recommender models, we rank a \its of items for every sequence in the test set using each model.
We calculate the metrics on the ranked items and then average the values for each model and rank the models using the mean.
In this paper we investigate three different strategies to create the \its of items and name the ranking according to the used method to extract the \its for calculating the metrics:
\begin{enumerate*}
    \item For the \textit{\exact ranking} we calculate the metrics on the target set that is equal to the full item set.
    \item The \its for the \textit{uniform ranking} consists of $\eta$ non-relevant items that are uniform random sampled from the item space and the relevant item~$i_r$.
    In this paper, we remove all items that are part of the sequence $\sequenceformodel$ from the sampled item space, because in our setting the user cannot interact with an item twice.
    So, the set of items, where we can sample from, for a sequence $\sequenceformodel$, is $N = \itemset \setminus \{i_r\} \setminus  \text{set}(\sequenceformodel)$.%, where $\text{set}(x)$ is a function that converts a sequence list to a set containing all items of sequence $x$.
    \item The \textit{popularity ranking} is the same as the previous ranking except we sample items from $N$ using the item's popularity.
\end{enumerate*}

Further, we adapt the definition of~\citeauthorwitcite{Krichene20} and define a ranking $R$ of a set of recommender models \textit{consistent} with ranking $Q$ of the same models iff the rank of all models in $R$ is equal to the rank in $Q$. 
That means, that the metric value of a recommender model in $R$ used to determine the ranking can be different to the one obtained for determining the rank in ranking $Q$, but two obtained ranks of the model must be the same.
\section{Related Work and Background}
\label{sec:relatedwork}

In this section we give a brief background overview about sequence recommendation and list the different random sampling evaluation settings used in previous model evaluation.
Further we list other studies that analyze the sampling evaluation methodology for recommender systems.

\subsection{Background}
In this subsection we provide some background about sequence recommender models and the evaluation methodologies used in previous work.

\subsubsection{Sequential Recommendation Models}
First work on recommender systems is based on Collaborative Filtering (CF) to model the user's interest given her previous interactions with items~\cite{sarwar2001itembased}.
All first proposed CF models, like Matrix Factorization~\cite{sarwar2001itembased}, ignore the order of the user's interactions to learn the user's preference.
To overcome this drawback, sequential recommendation models were introduced.
For example, \citeauthorwitcite{zimdars01} models the sequence for the recommendation task for the first time using first-order Markov Chains.
In recent years, models, that leverage different neural network types for encoding the sequence into a representation, have been published.
Early work used Recurrent Neural Networks (RNNs) with Gated Recurrent Units (GRUs)~\cite{cho2014learning} or a Long Short-Term Memory (LSTM)~\cite{hochreiter1997long} units as the encoder for the sequence.
While ~\citeauthorwitcite{hidasi2016sessionbased} use GRUs with a ranking loss, \citeauthorwitcite{tan2016improved} find that GRUs with cross-entropy loss outperforms GRUs without the ranking loss used in~\cite{hidasi2016sessionbased}.
A Convolutional Neural Network is used in the work of~\citeauthorwitcite{Tang18Caser}.
Attention, a mechanism that improves Natural Language Processing (NLP) tasks like machine translation~\cite{bahdanau2015neural,luong2015effective}, was first applied to the sequential recommendation task by~\citeauthorwitcite{li2017neural}.
Their model \narm uses attention with a GRU-based RNN network to encode the sequence into a representation.
In~\cite{kang2018selfattentive} the authors propose the \sasrec model, that is only based on attention and encodes the sequence using a unidirectional Transformer network~\cite{vaswani2017attention}.
\citeauthorwitcite{sun2019bert4rec} extend \sasrec by using a bidirectional encoder and adapt the BERT architecture and the training objective of the model, that is successfully used for language modelling in the NLP community, for the recommendation setting.
At the very time of this writing this model is the current state-of-the-art for sequence recommendation.%
\footnote{Here, we only consider models that exclusively use the sequence information and nothing else like, for example, the user or the time of the interaction.}

\subsubsection{Negative Sampling for Recommender Model Evaluation}
In this section, we will point to some work that uses random sampling to build the target set for evaluating recommender models.
\citeauthorwitcite{koren2008factorization} use for the first time random sampling to build target sets.
The author draws 1000 negative samples at random when evaluating different movie recommenders with the Root Mean Squared Error (RMSE) as metric. 
In~\cite{elkahky2015multiview} the authors use uniform negative random sampling with a size of 9 to evaluate their deep learning model for news, application and movie/tv recommendation using the Mean Reciprocal Rank (MRR) and \hrat{1}.
The current commonly used negative sample size of 100 samples was introduced by~\cite{he2017neural} to calculate the \hrat{10} and \ndcgat{10} for different recommenders on the \movielens dataset and a Pinterest dataset.
While the authors of the \sasrec model follow the procedure of~\cite{kang2018selfattentive} to calculate the Hit Rate and the NDCG, the authors of the \bertforrec model sample the 100 negative item based on the popularity distribution of the dataset~\cite{sun2019bert4rec}.
In contrast to this, in~\cite{huang2018improving} the authors apply a different approach and  draw 50 negative samples \textit{uniform} from the item set and another 50 items \textit{by their popularity}.

\subsection{Related Work}
Early work already studied the effects of the evaluation methodology while testing CF recommender methods~\cite{bellogin2011precisionoriented}.
In this work, the authors find that the ranking between the tested three matrix factorization models is inconsistent when evaluating them using four different methods, including \textit{uniform random sampling}, on the \movielenssystem \movielens dataset.
They calculate Recall, Precision and NDCG as metric to rank the studied models.
Later, \citeauthorwitcite{steck2013evaluation} also finds differences in model rankings of two matrix factorization models when using all unrated items or only the observed ratings in the test set of a user for a rating prediction setup and RMSE as metric.
In~\cite{bellogin2017statistical} the authors conduct studies varying the number of negative samples while comparing two collaborative filtering models, a k-Nearest Neighbor~(kNN)~\cite{cremonesi2010performance} and a probabilistic Latent Semantic Analysis~(pLSA)~\cite{hofmann2004latent} recommender model.
They find no ranking inconsistency in the rankings of the two models while testing on the \movielenssystem \movielens dataset obtained by metrics like Precision.
More recently, \citeauthorwitcite{Rendle19} shows on a constructed sample setting that evaluation using the \textit{uniform  sampling} is not consistent when using metrics like HR or NDCG.
In~\cite{Krichene20} the authors extend the work of~\citeauthor{Rendle19} by providing adapted metrics to overcome this problem.
They test the proposed corrected metrics on the \movielens dataset using a matrix factorization and two item-based collaborative filtering models~\cite{sarwar2001itembased} and find that the rankings stay consistent starting with fewer negative samples than without the corrected metrics.
At the same time, \citeauthor{Dong20} propose a dataset independent mapping function for the Hit Rate (HR), to approximate the HR obtained on the full item set of a model and therefore determine the full ranking of the considered models~\cite{Dong20}.
Their experiments conducted with collaborative filtering methods like MultiVAE~\cite{liang2018variational}, a variational autoencoder, for example, on the \movielens dataset demonstrate the applicability of their approximation function.
Closest to our work, \citeauthorwitcite{Canmares20} find that rankings of kNN~\cite{canamares2017probabilistic} and implicit matrix factorization~\cite{hu2008collaborative} on different sampling sizes drawn from a \textit{uniform} distribution using a k-fold evaluation on the \movielenssystem 1M dataset and the Yahoo! R3 dataset are inconsistent when varying the number of negative samples.

To the best of our knowledge this is the first work, that conduct studies about the consistency of recommender model rankings when evaluating the models by sampling negative items by popularity.
All previous work has only considered the uniform sampling approach in their analysis.
In addition, we compare the model rankings of current state-of-the-art sequential recommendation models, instead of simple CF recommendation models like~\cite{Canmares20,Rendle19,Dong20,Krichene20}, on different sampling based evaluation methods.
Furthermore, we perform these studies on five datasets commonly used for evaluation instead of just one or two datasets as before to find possible dependencies of rankings and evaluation methods on the dataset's properties.
We extend the work of~\cite{Canmares20}, that examine the consistency of the model rankings when uniform sampling different numbers of negative items, and use the provided method to perform the same investigations for sampling by popularity.
\section{Experimental Setup}
\label{sec:setup}

In this section we describe the experimental setup, including the recommender models and datasets we used for our analysis.
We also introduce the methods used to analyze the difference in the obtained model rankings.

\newcommand{\netinput}{\embeddingsize}
 
\subsection{Sequential Item Recommendation Models}
For investigating the random sampling evaluation, we use four state-of-the-art neural sequential recommendation networks.
Each network encodes the sequence $\sequence$ using three different sub-layers:
\begin{enumerate*}
    \item an embedding layer, that embeds each item $s_t$ in the sequence using an embedding matrix $M \in \mathbb{R}^{|\itemset| \times \embeddingsize}$, with $e_t = M_{s_t}$, where $M_j$ is the $j$-th row of matrix $M$,
    \item a sequence encoder, that transforms the embedded sequence to a representation $h \in \mathbb{R}^m$, where $m$ is the size of the representation,
    \item and an output layer, that projects the sequence representation back to the item space.
\end{enumerate*}
In the following we describe each used model in more detail.

\subsubsection{Gated Recurrent Unit (\gru)}
The \gru model of \citeauthorwitcite{tan2016improved} uses a Recurrent Neural Network (RNN) layer with Gated Recurrent Units (GRUs)~\cite{cho2014learning}, as the sequence encoder for the sequence.
RNNs encode sequences by learning internal hidden states $h_t \in \mathbb{R}^{\hiddensize}$, where $\hiddensize$ is the size of the sequence representation, at each sequence step $t$, given the current input $e_t$ and the hidden state of the previous step $h_{t-1}$.%
\footnote{We omit the output part of the RNN because we do not use this feature in our setting.}
Given $W_h \in \mathbb{R}^{\embeddingsize \times \hiddensize}$ and $R_h \in \mathbb{R}^{\hiddensize \times \hiddensize}$ as the weight matrices for the nonlinear transformation of the current input and the previous state and an activation function $g$, the hidden state of an RNN network is calculated by:
\begin{equation}
    h_t = g(W_h \netinput_t + R_h h_{t-1}).%
    \footnote{For this equation and all following equations in this paper we omit the bias term of the neural networks for readability.}
\end{equation}
RNNs suffer from the problem of vanishing gradients for long sequences.
To overcome this, the \gru memory unit utilizes two gates:
\begin{enumerate*}
    \item The \emph{update gate} $z$ controls how much of the previous state is passed to the next state(s) and
    \item the \emph{reset gate} $r$, that controls to which amount the network forgets the past state(s).
\end{enumerate*}
The state of recurrent network with GRUs can then be calculated with:
\begin{align}
    z_t &= \sigma(W_Z e_t + R_Z\ntimeprev{h}) \\
    r_t &= \sigma(W_R e_t + R_R\ntimeprev{h}) \\
    h_t &= (1 - z_t) \cdot \ntimeprev{h} + z_t \cdot \tanh(W_H e_t + R_H( r_t \cdot \ntimeprev{h})),
\end{align}
where $W_{Z, R, H} \in \mathbb{R}^{\hiddensize \times \embeddingsize}$ and $R_{Z, R, H} \in \mathbb{R}^{\hiddensize \times \hiddensize}$ are the weight matrices for the nonlinear transformation.
We use the last hidden state of the \gru $h_{l_{\sequence}}$ as the sequence representation and scale the network output to the item space using a feed forward network for the recommendation task: $o = W_s h_{l_{\sequence}}$ with $W_s \in \mathbb{R}^{n \times \hiddensize}$.
We train the model with all subsequences of all sequences and use the cross-entropy loss to optimize it:
\newcommand{\networkoutput}[3]{#1(o(#2))_{#3}}
\newcommand{\subsequence}[1]{\text{head}({\sequencesymbol}, #1)}
\begin{equation}
    L = -\sum_{s \in S} \sum_{t=1}^{s_{\sequencelength - 1}} \log( \networkoutput{\text{softmax}}{\subsequence{t}}{s_{t + 1}})
\end{equation}
\subsubsection{Neural Attentive Recommendation Machine (\narm)}
 The Neural Attentive Recommendation Machine (NARM)~\cite{li2017neural} consists of two different GRUs: %
 \begin{enumerate*}
     \item a global encoder and
     \item a local encoder
 \end{enumerate*}.
The global encoder is a GRU, that we already defined in the previous paragraph, thus the output of this encoder is $c^G = h_{\sequencelength}^G \in \mathbb{R}^{\hiddensize}$.
In contrast, the local encoder learns based on another GRU an attention to all hidden states of the sequence~\cite{bahdanau2015neural,luong2015effective}.
The output of the local encoder is defined as $c^L = \sum_{j = 1}^{\sequencelength} \alpha_{\sequencelength, j} h^L_j$, where $h^L_m \in \mathbb{R}^{\hiddensize}$ are the hidden states of the local encoder.
The attention score $\alpha_{\sequencelength,j}$ calculates an alignment between the j-th hidden state $\specifictime{h^L}{j}$ and the last hidden state $\ntime{h^L}$ using:
\begin{equation}
    \alpha_{\sequencelength,j} = q(\ntime{h^L}, \specifictime{h^L}{j})=V^\top \sigma(A_1 \ntime{h^L} + A_2 \specifictime{h^L}{j}),
\end{equation}
where, given a latent space $a$ for $\alpha$,\footnote{In practice the latent space is equal to the hidden space of the RNN encoder ($a = \hiddensize$).} $A_1$ and $A_2 \in \mathbb{R}^{\hiddensize \times a}$, $V \in \mathbb{R}^a$. % and $\sigma$ is the sigmoid activation function.
For the overall output of the encoder the representations of the global and local encoder are concatenated to $c = [c^G; c^L]$ and then transformed with a learn-able weight matrix $B \in \mathbb{R}^{\embeddingsize \times 2\hiddensize}$.
To scale the output to the item space, \narm uses the transposed item embedding matrix $M$: $o = B c M^\top$.
We train \narm the same way as the \gru model and also use the same loss.

\subsubsection{Self-Attention based Sequential Recommendation Model (\sasrec)}
The Self-Attention based Sequential Recommendation model (\sasrec)~\cite{kang2018selfattentive} is based on Transformer Networks~\cite{vaswani2017attention}.
Because Transformer Networks are not recurrent networks they do not have the information about the position, an additional embedding $P \in \mathbb{R}^{T \times \hiddensize}$, where $T$ is the maximum sequence length of the network and $d$ its embedding and sequence representation size, is added to learn the position $t$ of the item $s_t$ in the sequence: $\hat{e}_t = e_t + P_t$.
\sasrec consists of $L$ attention layer blocks, that are stacked on top of each other.
Each block at depth $j$ is a Multi-Head Self-Attention layer $\text{MH}^j$ followed by a pointwise feed-forward network $F$.
For a Multi-Head Self-Attention layer, a single Self-Attention layer $\text{SA}^j_m$ is applied $H$ times to its input, the results are concatenated and transformed using a weight matrix $W_O^j \in \mathbb{R}^{\hiddensize \times \hiddensize }$:
\begin{equation}
    \text{MH}^j(x) = [\text{SA}^j_1(x); \text{SA}^j_2(x); \ldots; \text{SA}^j_{H}(x)]W_O^j.
\end{equation}
Each Self-Attention layer linearly projects the input into a smaller space and then applies a mechanism called \textit{scaled-dot} attention~\cite{vaswani2017attention}:
\begin{equation}
    \text{SA}^j_m(x) = \text{attention}(xW_Q^j,xW_K^j,xW_V^j),
\end{equation}
where $W_Q^j,W_K^j, W_V^j \in \mathbb{R}^{d \times d/H}$ are linear projection matrices.
The attention function can be seen as a proportional retrieval of values $V$ given keys $K$ and queries $Q$ (in the case of self-attention $K=Q=V$) and is defined as:
\begin{equation}
    \text{attention}(Q, K, V) = \text{softmax} \left(\frac{QK^\top}{\sqrt{d}}\right) V,
\end{equation}
The pointwise feed-forward network F uses a ReLU~\cite{nair2010rectified} activation and two weight matrices $W^j_{F_1}$ and $W^j_{F_2} \in \mathbb{R}^{d \times d}$:
\begin{equation}
    F^j(x) = \text{ReLU}(MH^j(x)W^j_{F_1})W^j_{F_2}
\end{equation}
To stabilize training and to prevent overfitting, a residual connection~\cite{he2016residual} is added  to both layers $z$ of the block.
Further, layer normalization~\cite{ba2016layer} and Dropout~\cite{srivastava2014dropout} is applied to the input $x$ of the layers, resulting in:
\begin{equation}
    z(x) = x + \text{Dropout}(z(\text{LayerNormalization}(x)).
\end{equation}
\sasrec uses the state of the last sequence step $\hat{t}$ of the last layer $L$ as the representation of the sequence.
To project the presentation into the item space the model leverages the embedding matrix $M$ by $o(x) = F_{\hat{t}}^{L}(x) M^\top$.
At training time the network tries to predict each next sequence step in the sequence $\sequence_{\sequencelength}$ by taking $\sequence_{\sequencelength - 1}$ as input.
For this learning objective, all attention connection between $Q_m$ and $K_n$ with $m>n$ are masked to prevent the network to attend to subsequent items in the sequence.
\sasrec uses the BPR loss~\cite{rendle2009bayesian} for optimizing the network.
Therefore, it samples a negative item $i_{t, \text{neg}}$ (\ie an item that was not in the sequence and is not the target) randomly from the item set for each sequence step.
The overall loss is then:
\begin{equation}
    L = -\sum_{s \in S} \sum_{t=1}^{s_{\sequencelength - 1}} \log( \networkoutput{\sigma}{\subsequence{t}}{s_t+1}) + \log(1 - \networkoutput{\sigma}{\subsequence{t}}{i_{t, \text{neg}}}).
\end{equation}
\subsubsection{\bertforrec} The \bertforrec model~\cite{sun2019bert4rec} makes two modifications to the transformer architecture of \sasrec and one to the training objective.
In contrast to \sasrec, \bertforrec allows the model to attend to all positions of the sequence and uses GELU~\cite{hendrycks2016gaussian} instead of a ReLU as activation function for all non-linear feed-forward networks.
Instead of predicting the next item in the sequence, \bertforrec adapts the training objective of the language model BERT~\cite{devlin2019pretraining}, that uses the Cloze task~\cite{taylor1953cloze} for self-supervised training.
The model randomly masks items in the sequence and then tries to predict the masked item.
Since after training the prediction is generated by masking the last item in the sequence, the model additionally sometimes only masks the last item of the sequence while training.
Given the set $T_{s,\text{mask}}$, that contains the sequence positions of each masked item in the sequence $s$, the loss of the network is a cross entropy loss:
\begin{equation}
    L = - \sum_{s \in S} \sum_{t \in T_{s,\text{mask}}} \log(\networkoutput{\text{softmax}}{s}{s_t}).
\end{equation}

\subsection{Datasets}

\begin{table}[t]
    \caption{Statistics about the five datasets after pre-processing. We report the number of users, items and actions as well as the average sequence length (A. L.) and the density.}
    \label{tab:dataset-stats}
    \centering
    % generated by paper/datasets/build_dataset_table.py
\begin{tabular}{lrrrrr}
\toprule
Dataset & \#users & \#items & \#actions & Avg. Length & Density \\
\midrule
\ambeauty & \numprint{40226} & \numprint{54542} & \nprounddigits{1}\numprint{0.353962}m & \nprounddigits{2}\numprint{8.799333764232088} & \nprounddigits{2}\numprint{0.016133133666224358}\% \\
\amgame & \numprint{29341} & \numprint{23464} & \nprounddigits{1}\numprint{0.280945}m & \nprounddigits{2}\numprint{9.575167853856378} & \nprounddigits{2}\numprint{0.04080790936692967}\% \\
\movielens & \numprint{6040} & \numprint{3416} & \nprounddigits{1}\numprint{0.999611}m & \nprounddigits{2}\numprint{165.49850993377484} & \nprounddigits{2}\numprint{4.84480415496999}\% \\
\movielenslarge & \numprint{138493} & \numprint{26729} & \nprounddigits{1}\numprint{20.000263}m & \nprounddigits{2}\numprint{144.4135299257002} & \nprounddigits{2}\numprint{0.540287814455087}\% \\
\steam & \numprint{334537} & \numprint{13046} & \nprounddigits{1}\numprint{4.212176}m & \nprounddigits{2}\numprint{12.591061676286929} & \nprounddigits{2}\numprint{0.09651281370754966}\% \\
\bottomrule
\end{tabular}
\end{table}

We perform all experiments on five frequently used datasets that are based on user reviews in different domains.
The first two datasets \textit{\movielens} and \textit{\movielenslarge}\furl{https://grouplens.org/datasets/movielens/} are based on movie reviews that have been collected from the non-commercial movie recommendation site \movielenssystem.
While \movielens only contains about 1 million interactions, the \movielenslarge contains about 20 million interactions.
Further, we use datasets consisting of collected product reviews from Amazon organized into categories.%
\furl{http://jmcauley.ucsd.edu/data/amazon/links.html}
We will restrict our experiments to the two most widely used categories in related work: \textit{Beauty} and \textit{Games}.
The last dataset is comprised of game reviews crawled from the digital video game distribution service \steam.%
\furl{https://github.com/kang205/SASRec}

We apply the following common pre-processing steps: % to all datasets except \textit{ML-20m}.
First, all reviews are grouped by users and ordered by timestamp.
We treat a review of an item as an interaction with an item and building on that the sequence of interactions for each user.
Next, we remove all items and users with fewer than five occurrences from the dataset.
We follow the procedure in~\cite{sun2019bert4rec} and first build statistics for items and remove items under the threshold and then compute the user statistics and remove the users accordingly.
This is handled differently throughout the literature. For example, in \cite{kang2018selfattentive} statistics are computed on the original dataset and then both users and items are removed in one pass which leads to a slightly different dataset after pre-processing.
We skip the filtering step for \textit{ML-20m} since the statistics of the original dataset already closely match the reported statistics in different papers and application of the filtering yields large deviations from these reported values.
The overall statistics of all datasets after pre-processing can be found in \Cref{tab:dataset-stats}.

\subsection{Model Training and Verification}

In this section we describe the model training.
Further, we also verify that our implemented models achieve similar results to the previously reported results.%
\footnote{
All code for training and evaluating the models is available at \url{https://professor-x.de/papers/metrics-sampling-eval/}.
}
\subsubsection{Implementation and Training}
For the evaluation we implemented all four models using PyTorch.%
\footnote{https://pytorch.org}
Then, we retrain the three sequence models \sasrec and \bertforrec with the hyper-parameters reported in~\cite{sun2019bert4rec}, or if not reported we used the parameters found in the corresponding repositories of the papers.
For \gru and \narm we used the hyper-parameter of the corresponding papers.
All models are trained for a maximum of 800 epochs.
We use the Adam optimizer~\cite{kingma2015method} for training the networks and selected the best model based on the metric \hr{10} using the validation split.
All configurations of the models can be found in the additional material of this paper.
\begin{table}[t]
    \centering
    \caption{Performance comparison of our implementations of \gru, \sasrec and \bertforrec with values from \citeauthorwitcite{sun2019bert4rec} for the \hr{10} metric with popularity sampling and negative sample size $\eta = 100$. \gru is compared to the values of the \textit{GRU4Rec+}.}
     \begin{tabular}{lcccccc} % modified
    \toprule
    \multirow{2}{*}{\diagbox{Dataset}{Source}} & \multicolumn{2}{c}{\gru} & \multicolumn{2}{c}{\sasrec} & \multicolumn{2}{c}{\bertforrec} \\
    \cmidrule(l{2pt}r{2pt}){2-3} % added
    \cmidrule(l{2pt}r{2pt}){4-5} % added
    \cmidrule(l{2pt}r{2pt}){6-7} % added
 & ours & \cite{sun2019bert4rec} & ours & \cite{sun2019bert4rec} & ours & \cite{sun2019bert4rec} \\
    \midrule
    \ambeauty & \nprounddigits{3}\numprint{0.188} & \nprounddigits{3}\numprint{0.265} & \nprounddigits{3}\numprint{0.241} & \nprounddigits{3}\numprint{0.265} & \nprounddigits{3}\numprint{0.281} & \nprounddigits{3}\numprint{0.303} \\
    \movielens & \nprounddigits{3}\numprint{0.628} & \nprounddigits{3}\numprint{0.635} & \nprounddigits{3}\numprint{0.656} & \nprounddigits{3}\numprint{0.669} & \nprounddigits{3}\numprint{0.663} & \nprounddigits{3}\numprint{0.697} \\
    \movielenslarge & \nprounddigits{3}\numprint{0.651} & \nprounddigits{3}\numprint{0.652} & \nprounddigits{3}\numprint{0.706} & \nprounddigits{3}\numprint{0.714} & \nprounddigits{3}\numprint{0.732} & \nprounddigits{3}\numprint{0.747} \\
    \steam & \nprounddigits{3}\numprint{0.344} & \nprounddigits{3}\numprint{0.3594} & \nprounddigits{3}\numprint{0.384} & \nprounddigits{3}\numprint{0.378} & \nprounddigits{3}\numprint{0.414} & \nprounddigits{3}\numprint{0.401}\\
    \bottomrule
\end{tabular} 
    \label{tab:verify-model-results}
\end{table}

\subsubsection{Verifying Models}

To verify our implementations we compare the performance of our models with the reported ones in~\cite{sun2019bert4rec}.
Since \narm was not evaluated on the considered datasets in this paper, we can not verify our implementation.
But we will later show in the results section that the obtained relative rankings seems plausible.
Furthermore, we compare the results of our \gru implementation of~\cite{tan2016improved} with the results of \textit{GRU4Rec+}~\cite{hidasi2017topk}.
\Cref{tab:verify-model-results} shows our results for \hr{10} compared to the results reported in~\cite{sun2019bert4rec} with popularity sampling and $\eta=100$ negative samples.
Overall, the results of our implementations align with previously reported findings for this evaluation setting, where \bertforrec is best across the considered models and \sasrec comes in second, and \gru comes in third.
Our \gru implementation is competitive, compared with the reported values for \textit{GRU4Rec+} in \cite{sun2019bert4rec}.
Slight differences in the obtained metric scores, can be explained by small differences in parameter initialization due to the use of different neural network frameworks.
From these results we conclude that our implementations match the originals close enough for a meaningful comparison.

\subsection{Ranking Evaluation Methods}

To stabilize the obtained results, we run the experiments involving random sampling 20-times for each model.
We report the mean value over all 20 runs in the results section, but we omit the standard deviation, since we find that the standard deviation is zero, when we round it up the last three digits after the decimal point.
To compare two calculated rankings, we use the Kendall's Tau correlation coefficient~\cite{kendall1948rank} to determine how similar two rankings are.
Since we do not have ties in our rankings, we use the Kendall's Tau-a, that can be calculated by:
\begin{equation}
    \tau = \frac{2 (m_c - m_d)}{m (m-1)},
\end{equation}
where $m$ is the number of rankings (in our case 4) and $m_c$ the number of \textit{concordant} pairs and $m_d$ the number of \textit{discordant} pairs.
A pair $(x_1,y_1)$ and $(x_2, y_2)$ is \textit{concordant} if either $x_1 < x_2$ and $y_1 < y_2$ or $x_1 > x_2$ and $y_1 > y_2$, otherwise the pair is \textit{discordant}.
The values of the Kendall's Tau correlation range between $-1$ and $1$, where $-1$ indicates a perfect inversion and $1$ a perfect agreement of the rankings.

\section{Results}
\label{sec:results}

First, in this section we compare the \exact ranking of the considered models with the rankings achieved by sampling by popularity and uniform sampling.
Second, we consider how the number of negative samples affects the rankings and whether it can be chosen such that the rankings generated by the random selection strategies are consistent with the \exact ranking.

\begin{table*}[t]
    \centering
    \caption{The \exact, popular and uniform rankings of the four recommender models \gru, \narm, \sasrec and \bertforrec on the five considered datasets \ambeauty and Game, \movielenssystem \movielens and \movielenslarge  \steam with the corresponding \hr{10} metric, that was used to build the rankings.
    For both sample-based ranking we used $\eta = 100$ negative samples and we also report the \kendallc $\tau$ with respect to the \exact ranking in the first column.
    For each dataset the best \hr{10} (highest) value is marked as bold and second best value is underlined.
    Additionally, the ranking is visualized with $\blacksquare$ symbols for each dataset independently.
    The number of $\blacksquare$ symbols corresponds to the achieved rank (fewer symbols are better).
    }

    \begin{tabular}{llrlrlcrlc}
    \toprule
 {} & {}      & \multicolumn{2}{c}{\exact} & \multicolumn{3}{c}{popularity} & \multicolumn{3}{c}{uniform} \\
     \cmidrule(l{2pt}r{2pt}){3-4} % added
    \cmidrule(l{2pt}r{2pt}){5-7} % added
    \cmidrule(l{2pt}r{2pt}){8-10} % added
 Dataset   & Model   & \hrat{10}  & rank           & \hrat{10}  & rank  & $\tau$        & \hrat{10}  & rank & $\tau$ \\
    \midrule
\multirow{4}{*}{\ambeauty}  &     \gru  & \nprounddigits{3}\numprint{0.0312733054161071}  & $\blacksquare\blacksquare\blacksquare$  & \nprounddigits{3}\numprint{0.18844652846455573}  & $\blacksquare\blacksquare\blacksquare\blacksquare$  & \multirow{4}{*}{\nprounddigits{2}\numprint{-0.3333333333333334}}  & \nprounddigits{3}\numprint{0.3494443893432617}  & $\blacksquare\blacksquare\blacksquare\blacksquare$  & \multirow{4}{*}{\nprounddigits{2}\numprint{0.0}} \\
 &     \narm  & \underline{\nprounddigits{3}\numprint{0.03261572122573851}}  & $\blacksquare\blacksquare$  & \underline{\nprounddigits{3}\numprint{0.2426428191363811}}  & $\blacksquare\blacksquare$  &   & \nprounddigits{3}\numprint{0.41680753976106644}  & $\blacksquare\blacksquare\blacksquare$  &  \\
 &     \sasrec  & \textbf{\nprounddigits{3}\numprint{0.03559886664152141}}  & $\blacksquare$  & \nprounddigits{3}\numprint{0.24114503189921374}  & $\blacksquare\blacksquare\blacksquare$  &   & \underline{\nprounddigits{3}\numprint{0.4196527123451233}}  & $\blacksquare\blacksquare$  &  \\
 &     \bertforrec  & \nprounddigits{3}\numprint{0.027345497161149906}  & $\blacksquare\blacksquare\blacksquare\blacksquare$  & \textbf{\nprounddigits{3}\numprint{0.28060831129550934}}  & $\blacksquare$  &   & \textbf{\nprounddigits{3}\numprint{0.4250857636332512}}  & $\blacksquare$  &  \\
\hline
\multirow{4}{*}{\amgame}  &     \gru  & \nprounddigits{3}\numprint{0.0661531612277031}  & $\blacksquare\blacksquare\blacksquare\blacksquare$  & \nprounddigits{3}\numprint{0.38028867542743683}  & $\blacksquare\blacksquare\blacksquare\blacksquare$  & \multirow{4}{*}{\nprounddigits{2}\numprint{0.6666666666666669}}  & \nprounddigits{3}\numprint{0.5960550099611283}  & $\blacksquare\blacksquare\blacksquare\blacksquare$  & \multirow{4}{*}{\nprounddigits{2}\numprint{0.6666666666666669}} \\
 &     \narm  & \textbf{\nprounddigits{3}\numprint{0.08667053282260893}}  & $\blacksquare$  & \underline{\nprounddigits{3}\numprint{0.5139974772930145}}  & $\blacksquare\blacksquare$  &   & \underline{\nprounddigits{3}\numprint{0.7115350544452668}}  & $\blacksquare\blacksquare$  &  \\
 &     \sasrec  & \nprounddigits{3}\numprint{0.06755052506923673}  & $\blacksquare\blacksquare\blacksquare$  & \nprounddigits{3}\numprint{0.4512473985552788}  & $\blacksquare\blacksquare\blacksquare$  &   & \nprounddigits{3}\numprint{0.6881616234779357}  & $\blacksquare\blacksquare\blacksquare$  &  \\
 &     \bertforrec  & \underline{\nprounddigits{3}\numprint{0.08166047185659402}}  & $\blacksquare\blacksquare$  & \textbf{\nprounddigits{3}\numprint{0.5184298425912857}}  & $\blacksquare$  &   & \textbf{\nprounddigits{3}\numprint{0.7195358037948608}}  & $\blacksquare$  &  \\
\hline
\multirow{4}{*}{\movielens}  &     \gru  & \textbf{\nprounddigits{3}\numprint{0.2240066230297088}}  & $\blacksquare$  & \nprounddigits{3}\numprint{0.6276655673980713}  & $\blacksquare\blacksquare\blacksquare$  & \multirow{4}{*}{\nprounddigits{2}\numprint{-0.6666666666666669}}  & \underline{\nprounddigits{3}\numprint{0.7675082713365555}}  & $\blacksquare\blacksquare$  & \multirow{4}{*}{\nprounddigits{2}\numprint{0.3333333333333334}} \\
 &     \narm  & \underline{\nprounddigits{3}\numprint{0.2023178786039352}}  & $\blacksquare\blacksquare$  & \nprounddigits{3}\numprint{0.6275082767009735}  & $\blacksquare\blacksquare\blacksquare\blacksquare$  &   & \nprounddigits{3}\numprint{0.7637500017881393}  & $\blacksquare\blacksquare\blacksquare$  &  \\
 &     \sasrec  & \nprounddigits{3}\numprint{0.18460264801979062}  & $\blacksquare\blacksquare\blacksquare$  & \underline{\nprounddigits{3}\numprint{0.6556456923484802}}  & $\blacksquare\blacksquare$  &   & \textbf{\nprounddigits{3}\numprint{0.7983443737030029}}  & $\blacksquare$  &  \\
 &     \bertforrec  & \nprounddigits{3}\numprint{0.15993377566337583}  & $\blacksquare\blacksquare\blacksquare\blacksquare$  & \textbf{\nprounddigits{3}\numprint{0.6625165581703186}}  & $\blacksquare$  &   & \nprounddigits{3}\numprint{0.7628642380237579}  & $\blacksquare\blacksquare\blacksquare\blacksquare$  &  \\
\hline
\multirow{4}{*}{\movielenslarge}  &     \gru  & \underline{\nprounddigits{3}\numprint{0.1992519497871399}}  & $\blacksquare\blacksquare$  & \nprounddigits{3}\numprint{0.651153489947319}  & $\blacksquare\blacksquare\blacksquare$  & \multirow{4}{*}{\nprounddigits{2}\numprint{0.6666666666666669}}  & \underline{\nprounddigits{3}\numprint{0.9638559371232986}}  & $\blacksquare\blacksquare$  & \multirow{4}{*}{\nprounddigits{2}\numprint{0.0}} \\
 &     \narm  & \nprounddigits{3}\numprint{0.11838143318891522}  & $\blacksquare\blacksquare\blacksquare\blacksquare$  & \nprounddigits{3}\numprint{0.568571338057518}  & $\blacksquare\blacksquare\blacksquare\blacksquare$  &   & \nprounddigits{3}\numprint{0.9521203219890595}  & $\blacksquare\blacksquare\blacksquare\blacksquare$  &  \\
 &     \sasrec  & \nprounddigits{3}\numprint{0.13680113852024073}  & $\blacksquare\blacksquare\blacksquare$  & \underline{\nprounddigits{3}\numprint{0.7062367826700211}}  & $\blacksquare\blacksquare$  &   & \textbf{\nprounddigits{3}\numprint{0.9737463295459747}}  & $\blacksquare$  &  \\
 &     \bertforrec  & \textbf{\nprounddigits{3}\numprint{0.24109521508216852}}  & $\blacksquare$  & \textbf{\nprounddigits{3}\numprint{0.732035544514656}}  & $\blacksquare$  &   & \nprounddigits{3}\numprint{0.9638162195682526}  & $\blacksquare\blacksquare\blacksquare$  &  \\
\hline
\multirow{4}{*}{\steam}  &     \gru  & \underline{\nprounddigits{3}\numprint{0.19785255193710324}}  & $\blacksquare\blacksquare$  & \nprounddigits{3}\numprint{0.343864506483078}  & $\blacksquare\blacksquare\blacksquare\blacksquare$  & \multirow{4}{*}{\nprounddigits{2}\numprint{0.0}}  & \nprounddigits{3}\numprint{0.8289144426584244}  & $\blacksquare\blacksquare\blacksquare$  & \multirow{4}{*}{\nprounddigits{2}\numprint{0.6666666666666669}} \\
 &     \narm  & \nprounddigits{3}\numprint{0.1963549554347992}  & $\blacksquare\blacksquare\blacksquare$  & \nprounddigits{3}\numprint{0.3552278220653534}  & $\blacksquare\blacksquare\blacksquare$  &   & \underline{\nprounddigits{3}\numprint{0.8403080970048904}}  & $\blacksquare\blacksquare$  &  \\
 &     \sasrec  & \nprounddigits{3}\numprint{0.1829423904418945}  & $\blacksquare\blacksquare\blacksquare\blacksquare$  & \underline{\nprounddigits{3}\numprint{0.38370389491319656}}  & $\blacksquare\blacksquare$  &   & \nprounddigits{3}\numprint{0.8263523876667023}  & $\blacksquare\blacksquare\blacksquare\blacksquare$  &  \\
 &     \bertforrec  & \textbf{\nprounddigits{3}\numprint{0.21466384828090662}}  & $\blacksquare$  & \textbf{\nprounddigits{3}\numprint{0.4142106890678406}}  & $\blacksquare$  &   & \textbf{\nprounddigits{3}\numprint{0.8605187147855758}}  & $\blacksquare$  &  \\
\bottomrule
    \end{tabular}

    \label{tab:true-ranking-recall}
\end{table*}

\newcommand{\figureoverallsize}{0.85}
\newcommand{\subfiguresize}{0.22}
% rank vs sampling strategy plots
\begin{figure}
\begin{subfigure}{\subfiguresize\textwidth}
  \centering
  % include first image
  \includegraphics[width=\figureoverallsize\linewidth]{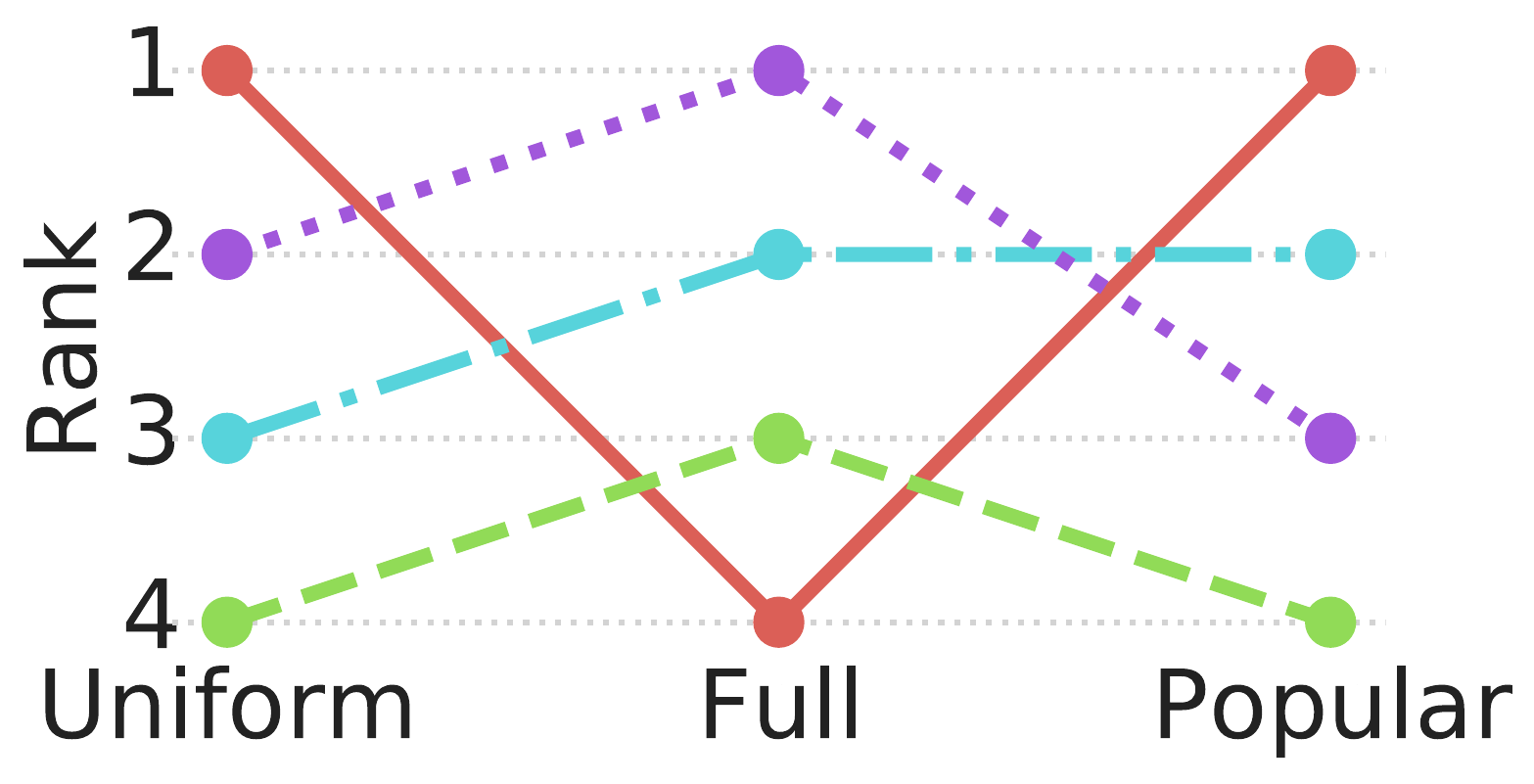}  
  \caption{\ambeauty}
  \label{fig:ranks-strategy-beauty}
\end{subfigure}
\hfill
\begin{subfigure}{\subfiguresize\textwidth}
  \centering
  % include second image 
  \includegraphics[width=\figureoverallsize\linewidth]{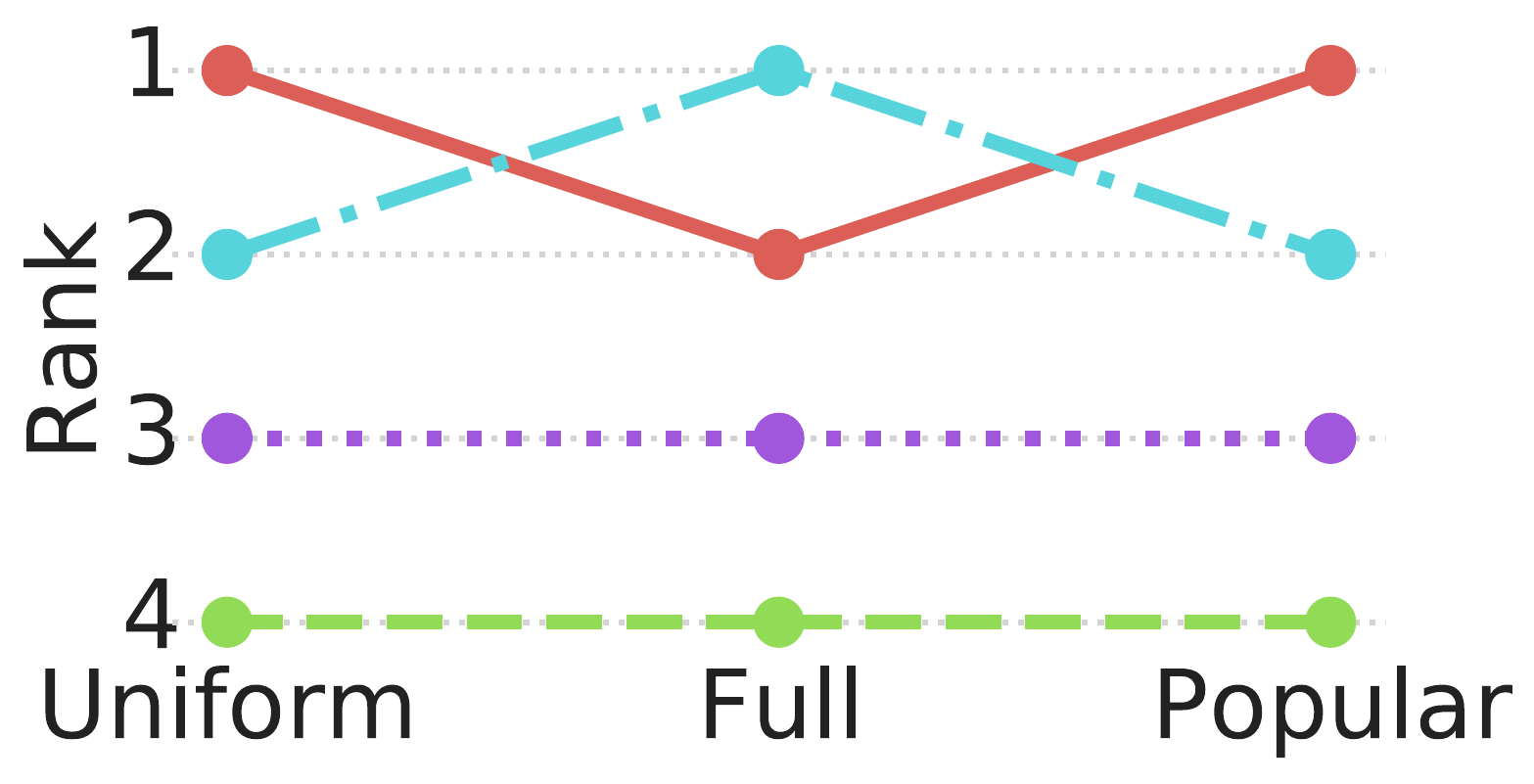}  
  \caption{\amgame}
  \label{fig:ranks-strategy-games}
\end{subfigure}
\hfill
\begin{subfigure}{\subfiguresize\textwidth}
  \centering
  % include third image
  \includegraphics[width=\figureoverallsize\linewidth]{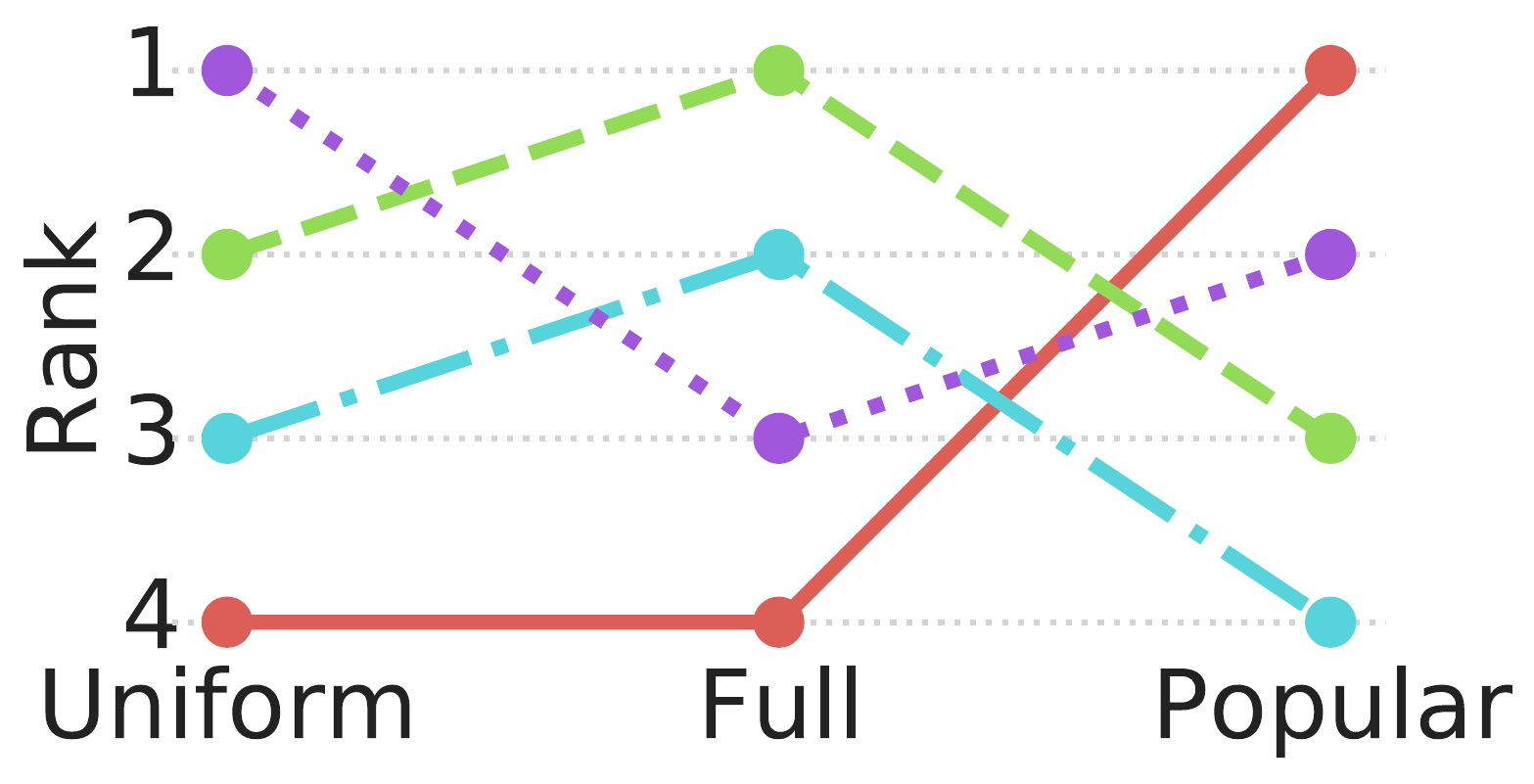}
  \caption{\movielens}
  \label{fig:ranks-strategy-movielens}
\end{subfigure}
\hfill
\begin{subfigure}{\subfiguresize\textwidth}
  \centering
  % include fourth image
  \includegraphics[width=\figureoverallsize\linewidth]{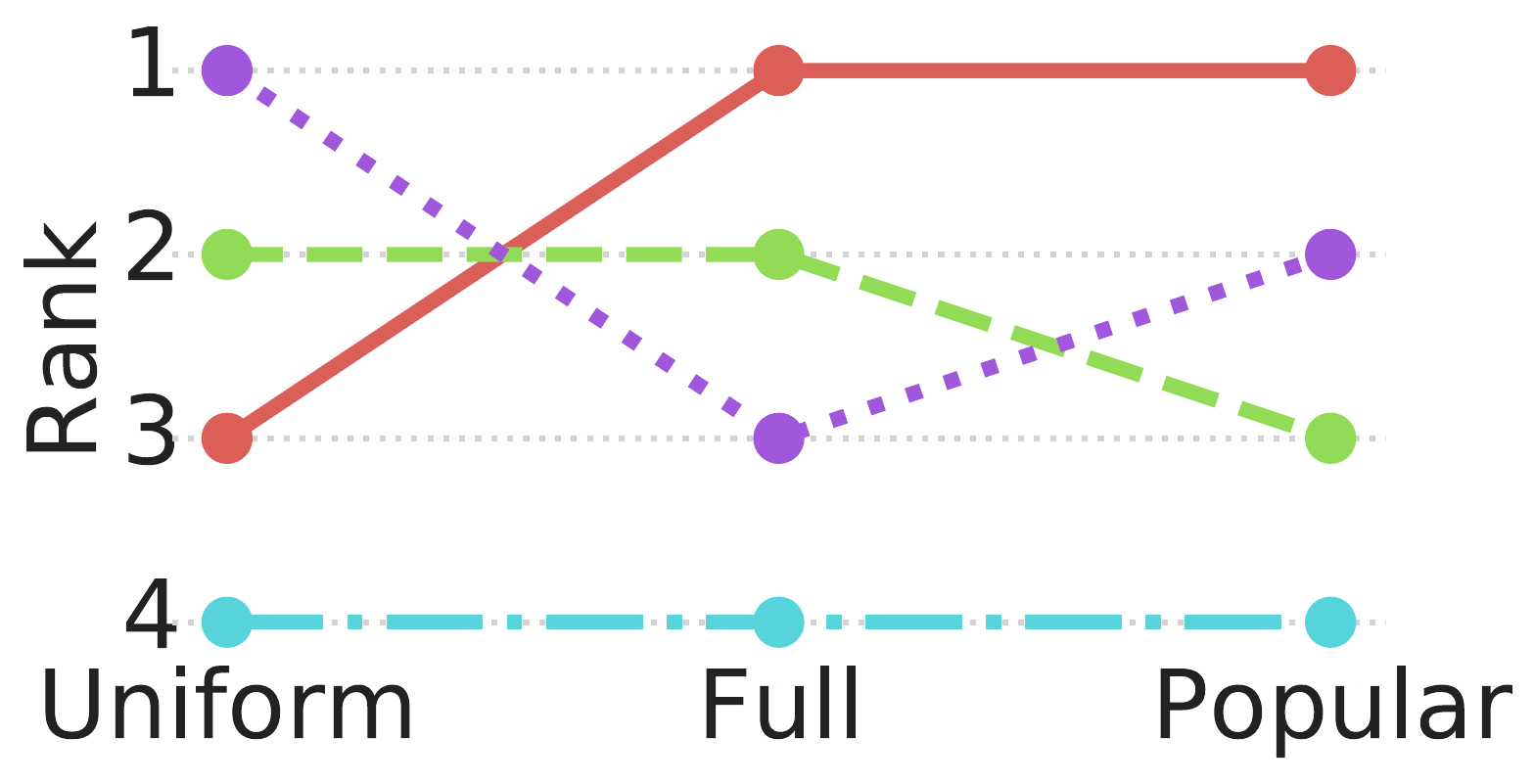}  
  \caption{\movielenslarge}
  \label{fig:ranks-strategy-movielenslarge}
\end{subfigure}
\hfill
\begin{subfigure}{\subfiguresize\textwidth}
  \centering
  % include third image
  \includegraphics[width=\figureoverallsize\linewidth]{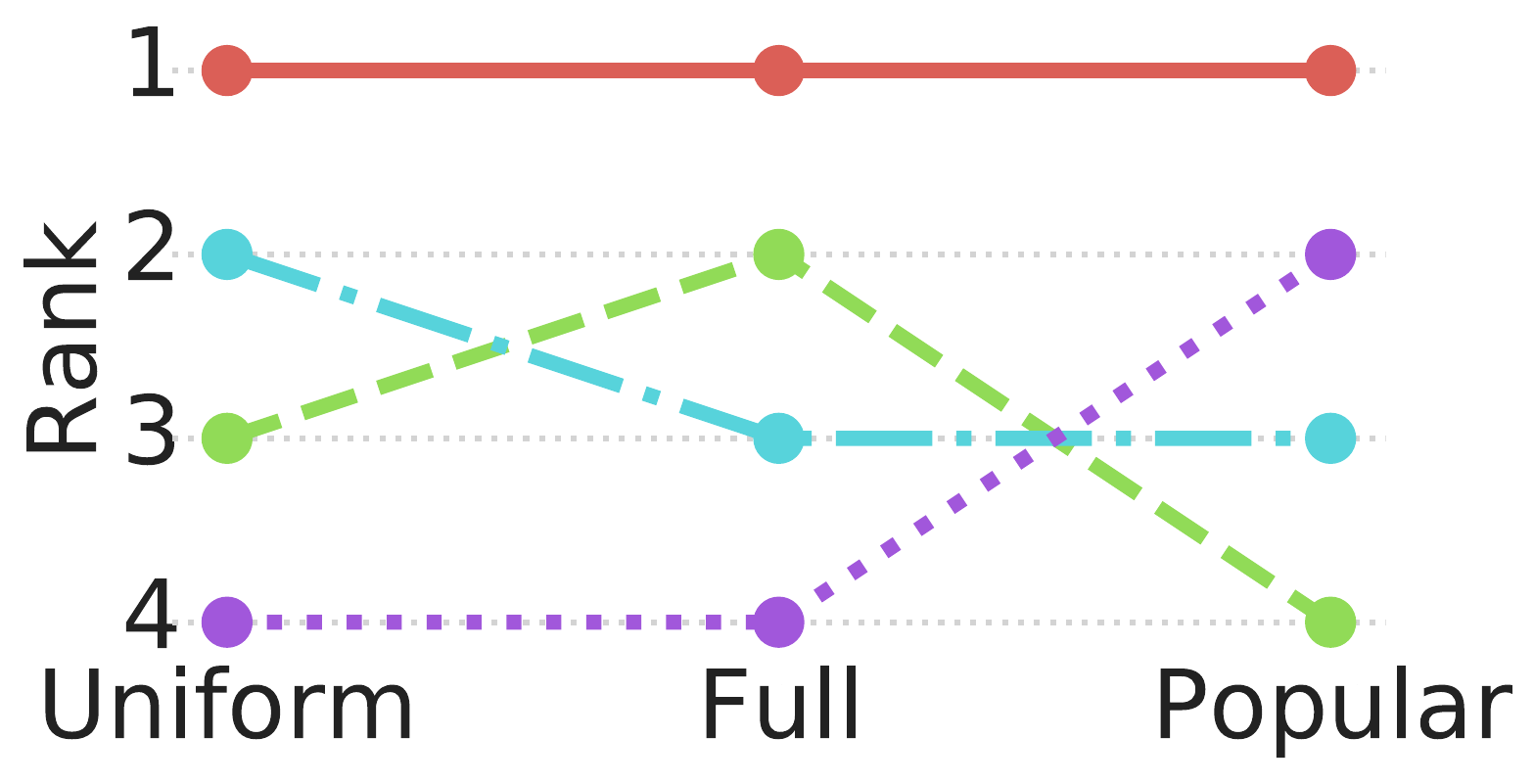}  
  \caption{\steam}
  \label{fig:ranks-strategy-steam}
\end{subfigure}
\hfill
\begin{subfigure}[m]{\subfiguresize\textwidth}
  \centering
  % include third image
  \includegraphics[width=\linewidth]{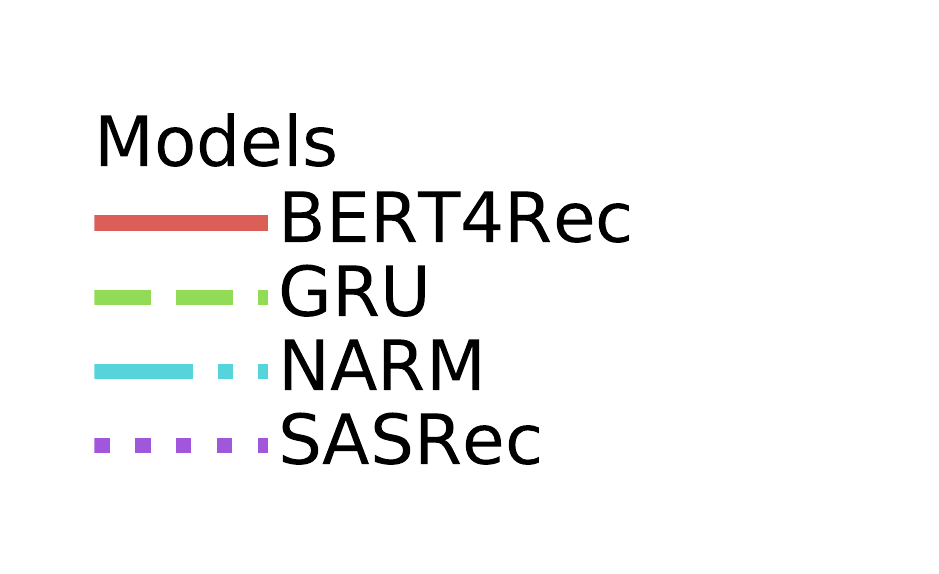}  
\end{subfigure}
\Description[Comparison of relative model ranking]{Comparison of relative model ranking for \hrate{10} with sampling strategies and \exact ranking.}
\caption{Rankings based on \hrat{10} of the four recommender models \gru, \narm, \sasrec and \bertforrec on the five considered datasets \ambeauty and Game, \movielenssystem \movielens and \movielenslarge and \steam.
For each dataset we compare the the \textit{\exact ranking} in the middle with \textit{uniform ranking} on the left side and the \textit{popular ranking} on the right side.
The number of negative samples is set to $\eta = 100$ for both sampling rankings.}
\label{fig:rankings-hr10}
\end{figure}

\subsection{\Exact Ranking and Random Sampling Rankings}
In this section we take a look at how well the rankings based on one of the two sampling strategies approximates the \exact ranking.
Therefore, we evaluate the model rankings with all three options (\exact, uniform and popular) for both metrics \hrat{10} and \ndcg{10}. % to also include a rank sensitive metric in the analysis.
Additionally, we calculate the \kendall correlation ($\tau$) between the sampling rankings and the \exact ranking.
For the uniform and popular rankings, we use $\eta = 100$ negative samples which is used by recent work~\cite{kang2018selfattentive, sun2019bert4rec}.

\Cref{tab:true-ranking-recall} shows the obtained three rankings of the four models on the five datasets with the corresponding metric scores for \hrat{10} and the corresponding \kendall correlation values between the sampled rankings and the \exact ranking. % the model ranking obtained through uniform random sampling and sampling by popularity in comparison with the \exact ranking when measuring the model performance using the \hr{10} metric including the Kendall-Tau correlation  ($\tau$) between the sampling evaluation methods and the \exact ranking.
\Cref{fig:rankings-hr10} visualizes the difference of the three model rankings for each dataset.
As we can observe from the table, the uniform rankings are not consistent with the \exact ranking across all five datasets when comparing the four models.
For example, the inconsistency is especially evident when considering the \kendallc for the \ambeauty dataset, where $\tau = \nprounddigits{2}\numprint{0}$, that is, that there are no matches in rank between the two rankings.
This can also clearly be seen in \Cref{fig:ranks-strategy-beauty}.
Overall, our findings are in line with the ones by \citeauthorwitcite{Krichene20}.
Furthermore, from this it can be concluded that the results in~\cite{Krichene20} obtained on a matrix factorization and two collaborative filtering models from~\cite{sarwar2001itembased} can be extended to the more recent state-of-the-art deep learning models.
When we now look at the popularity ranking, we find that this ranking is also not consistent with the \exact ranking like the uniform ranking.
Indeed for \movielens we report $\tau = \nprounddigits{2}\numprint{-0.67}$, which means that the ranking is almost inverse (see \Cref{fig:ranks-strategy-movielens}) to the \exact ranking.
After all (when we exclude \narm, since it was not evaluated on the five considered datasets) we can confirm the model ranking \gru, \sasrec, \bertforrec on the datasets \ambeauty, \movielens, \movielenslarge and \steam reported by~\cite{sun2019bert4rec} using the popular ranking.
But looking at the full ranking, we can see that the models perform differently well on the five datasets.
For example, when considering the \exact ranking the \bertforrec model outperforms the other methods on the \movielenslarge and \steam datasets, but is the model with the worst \hrat{10} on the \movielens and \ambeauty datasets.
This fact suggests that the model ranking depends on the evaluation method used (with or without sampling).
At last, when we compare both sampling methods with each other, \Cref{tab:true-ranking-recall} and \Cref{fig:rankings-hr10} reveal that both sampling methods are mostly inconsistent with each other.
The exception is the \amgame dataset, where both sampling methods agree on a ranking, but are still inconsistent with the \exact ranking.
Due to space limitations, we omit the table for the rankings using \ndcg{10} as metric, since we observe overall the same behavior on all rankings on all datasets when using \ndcg{10} instead of \hrat{10}.

To summarize, with our findings we extend previous results, that uniform rankings are inconsistent with the rankings obtained on the full item set, to recent state-of-the-art sequential recommendation methods.
Furthermore, we find that the popular rankings are also inconsistent with the exact ranking when considering 100 negative samples ($\eta = 100$).

\newcommand{\samplefiguresize}{0.35}
% rank vs. sampling size plots
\begin{figure*}
\begin{subfigure}{\textwidth}
  \centering
  % include first image
  \includegraphics[width=\samplefiguresize\linewidth]{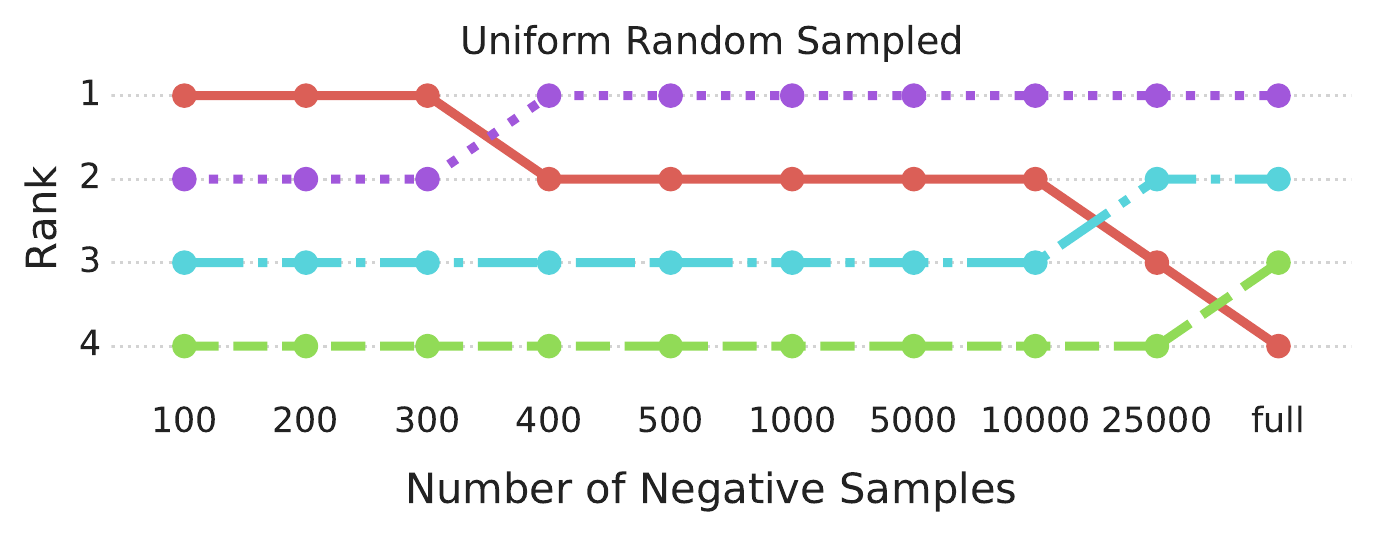}
  \includegraphics[width=\samplefiguresize\linewidth]{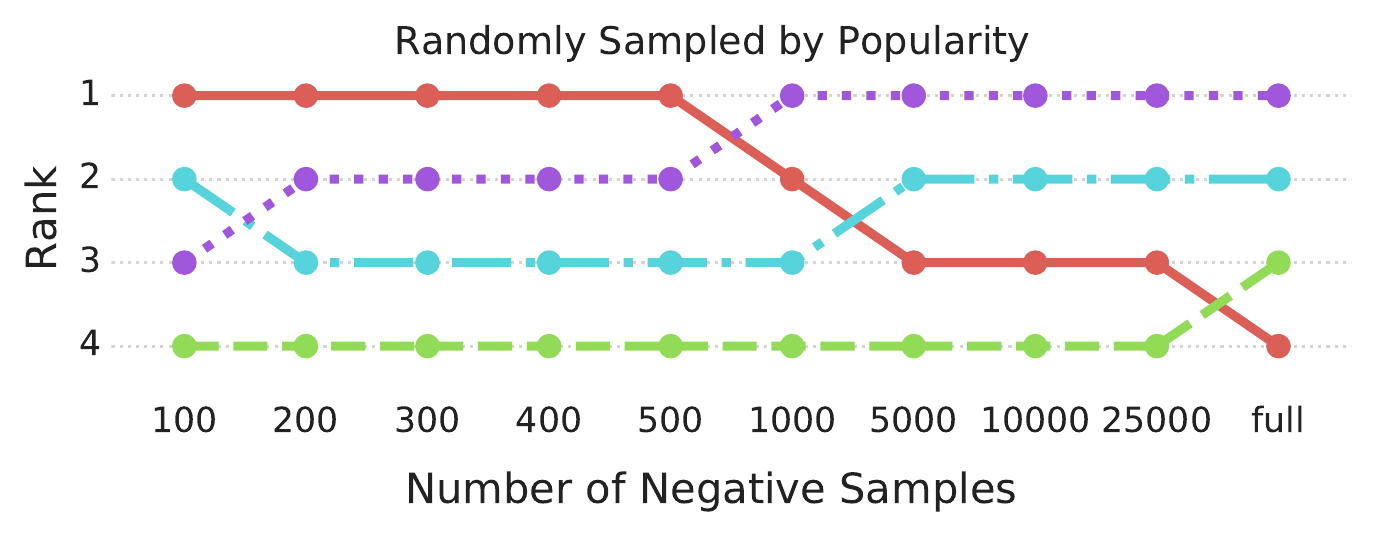}
  \caption{\ambeauty}
  \label{fig:ranks-by-size-beauty}
\end{subfigure}

\begin{subfigure}{\textwidth}
  \centering
  % include first image
  \includegraphics[width=\samplefiguresize\linewidth]{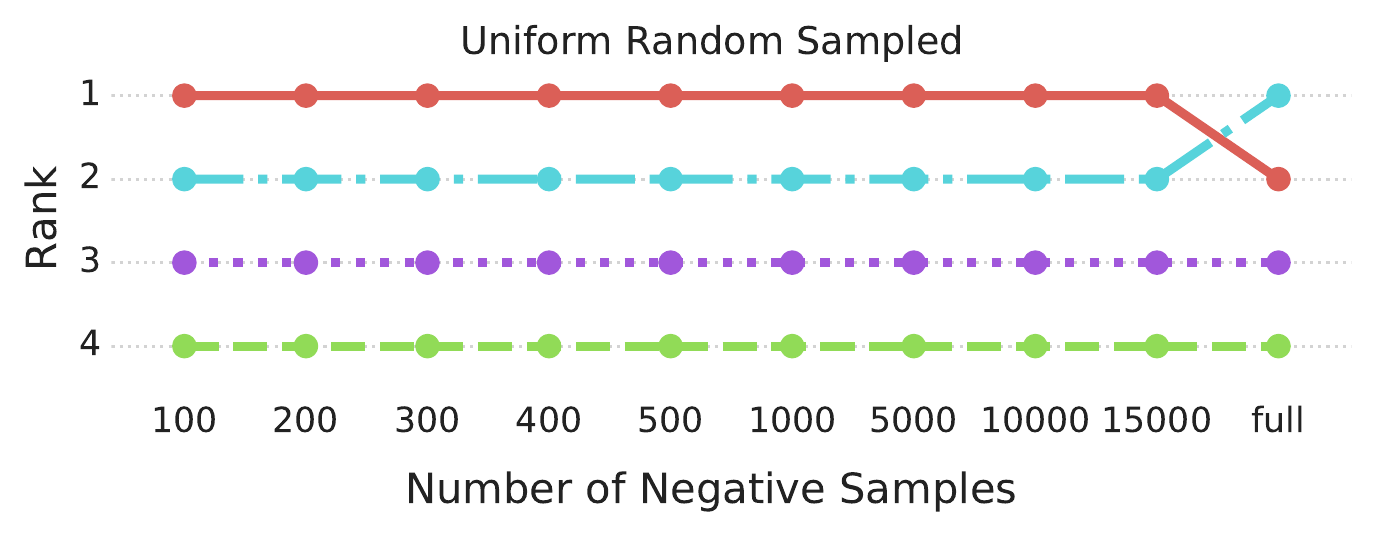}  
  \includegraphics[width=\samplefiguresize\linewidth]{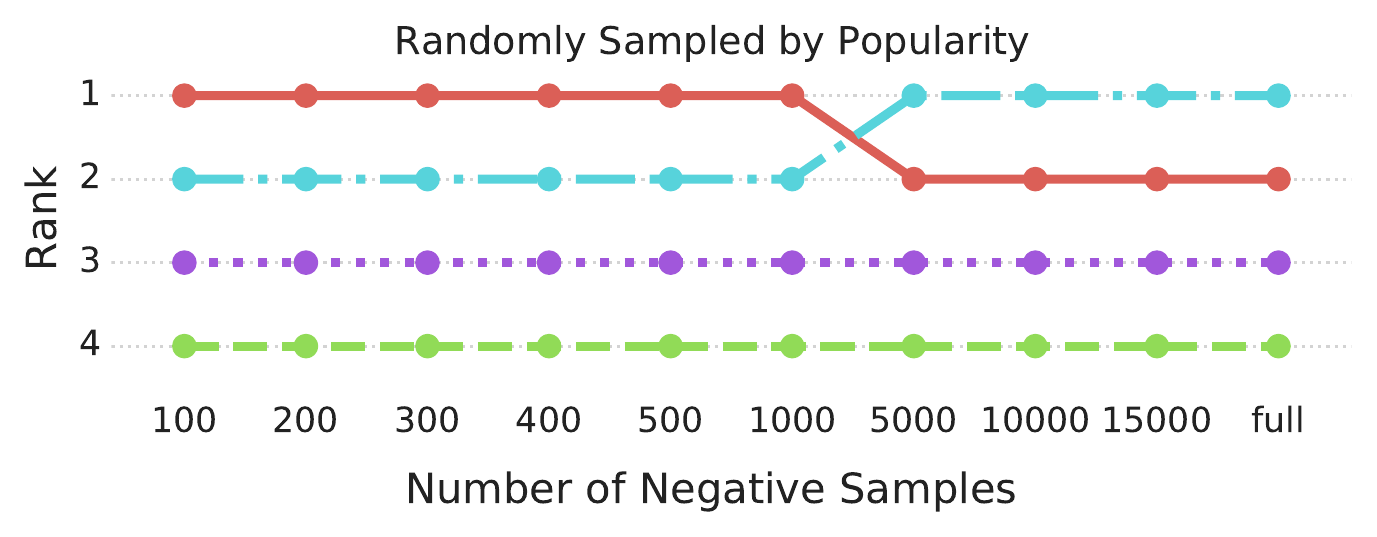} 
  \caption{\amgame}
  \label{fig:ranks-by-size-games}
\end{subfigure}

\begin{subfigure}{\textwidth}
  \centering
  % include first image
  \includegraphics[width=\samplefiguresize\linewidth]{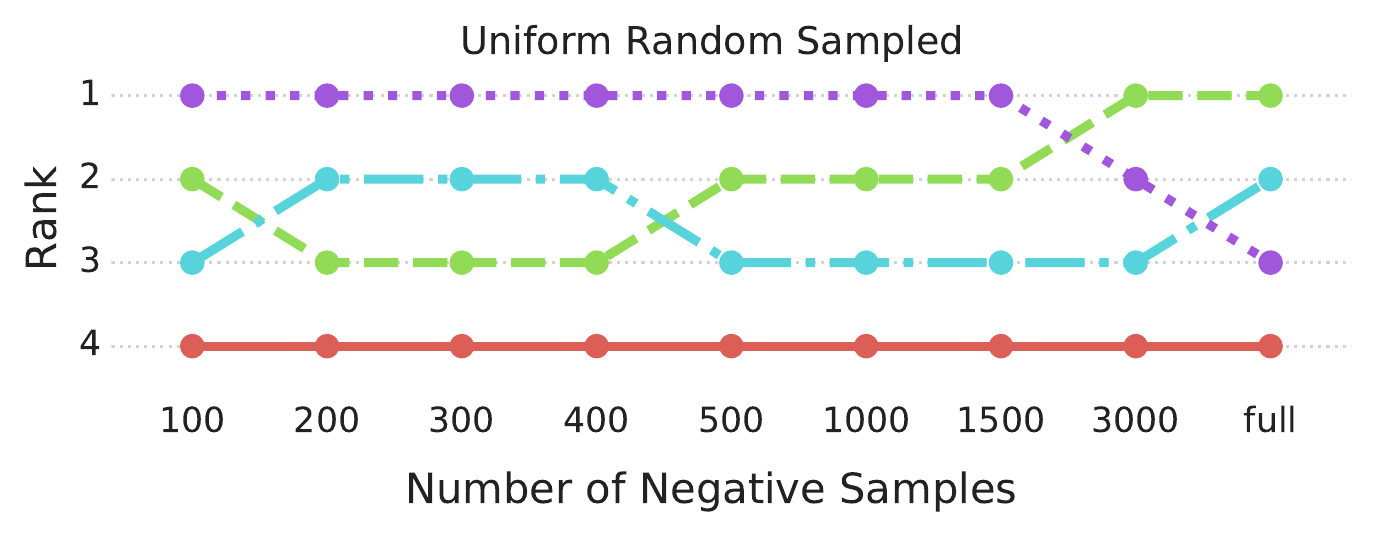}
  \includegraphics[width=\samplefiguresize\linewidth]{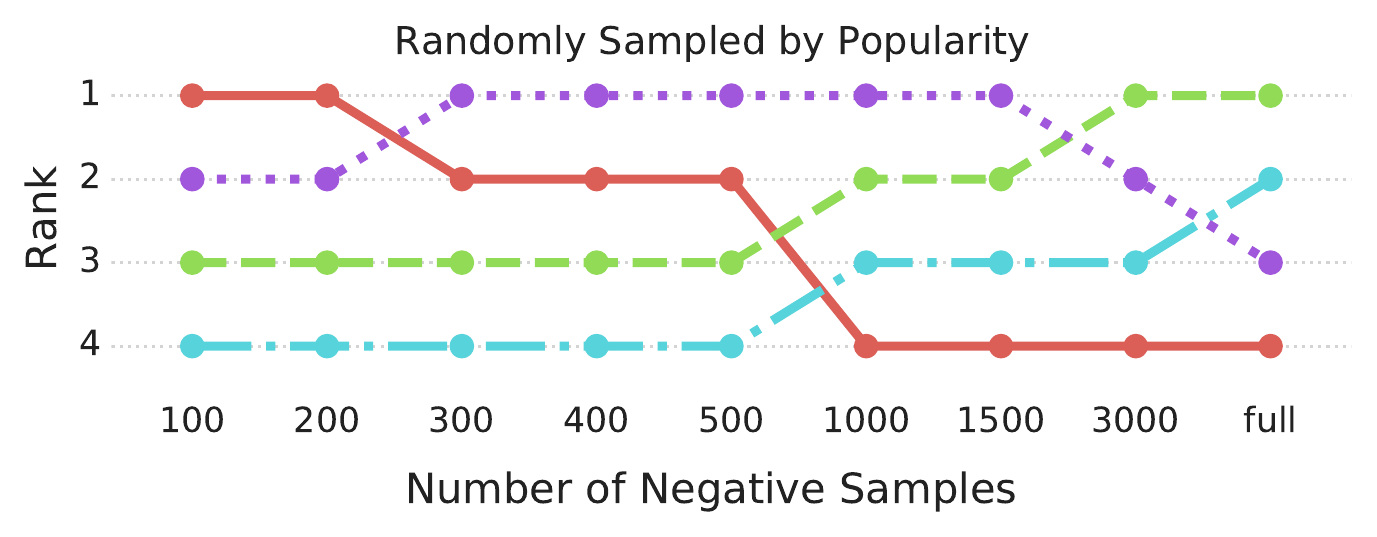}  
  \caption{\movielens}
  \label{fig:ranks-by-size-ml1m}
\end{subfigure}

\begin{subfigure}{\textwidth}
  \centering
  % include first image
  \includegraphics[width=\samplefiguresize\linewidth]{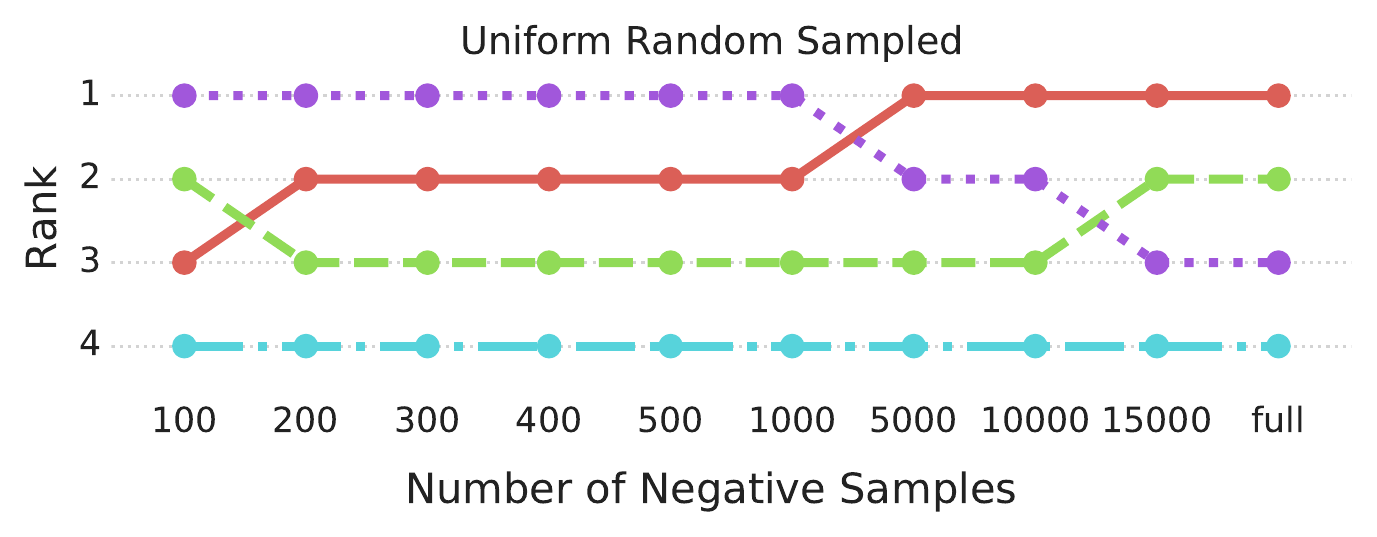}  
  \includegraphics[width=\samplefiguresize\linewidth]{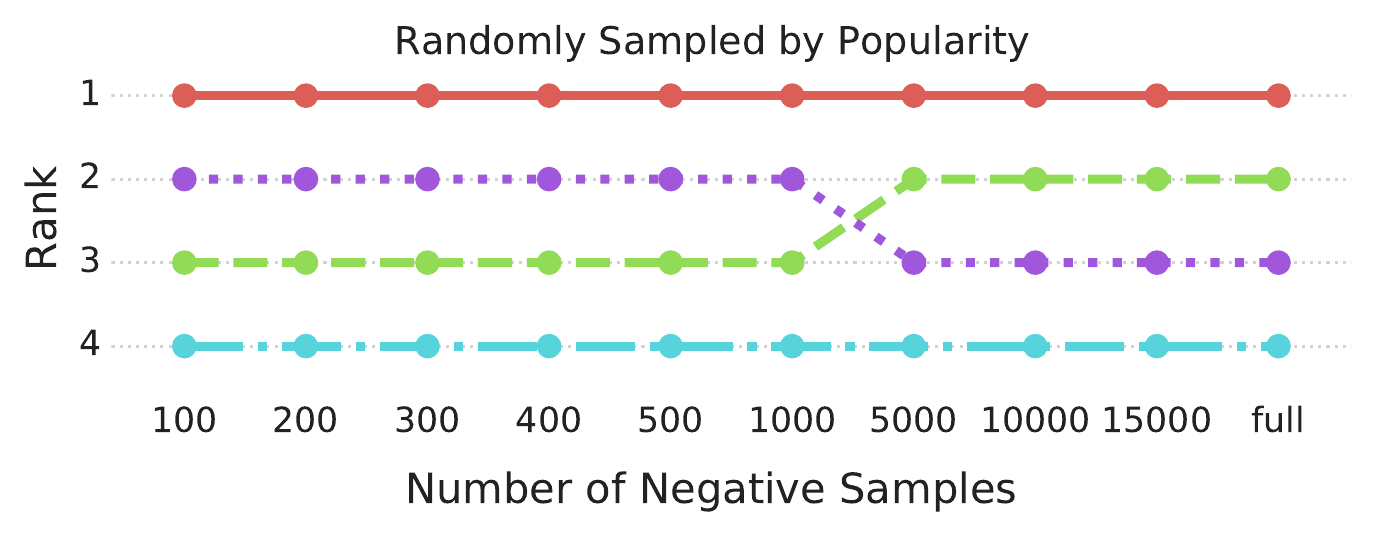}
  \caption{\movielenslarge}
  \label{fig:ranks-by-size-ml20m}
\end{subfigure}

\begin{subfigure}{\textwidth}
  \centering
  % include first image
  \includegraphics[width=\samplefiguresize\linewidth]{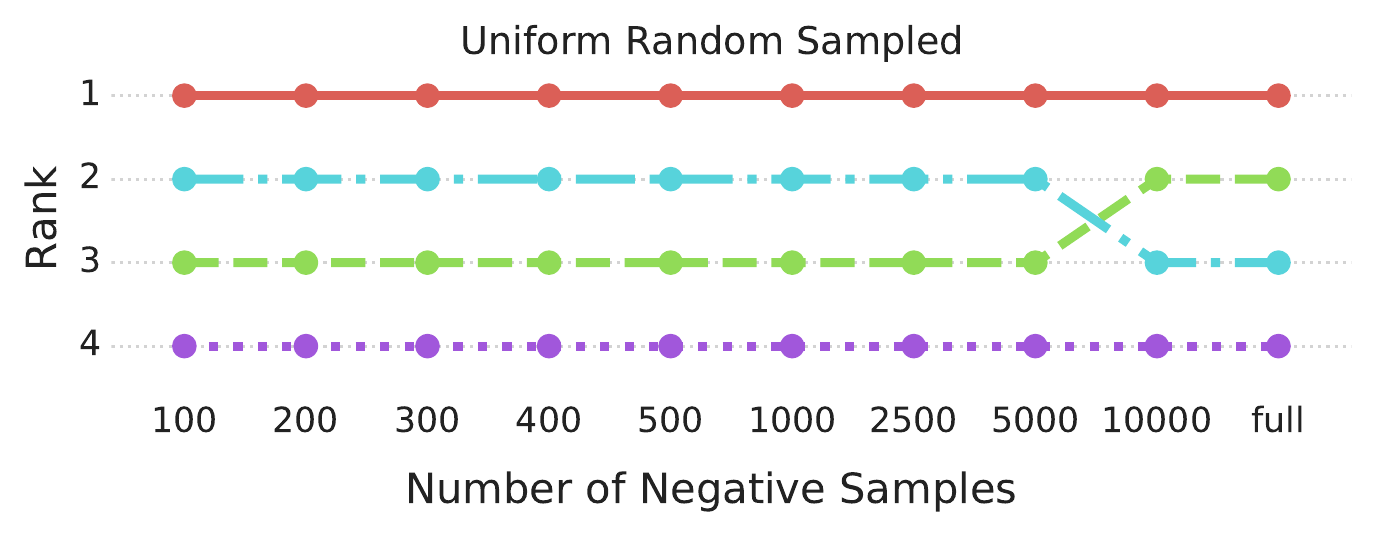}
  \includegraphics[width=\samplefiguresize\linewidth]{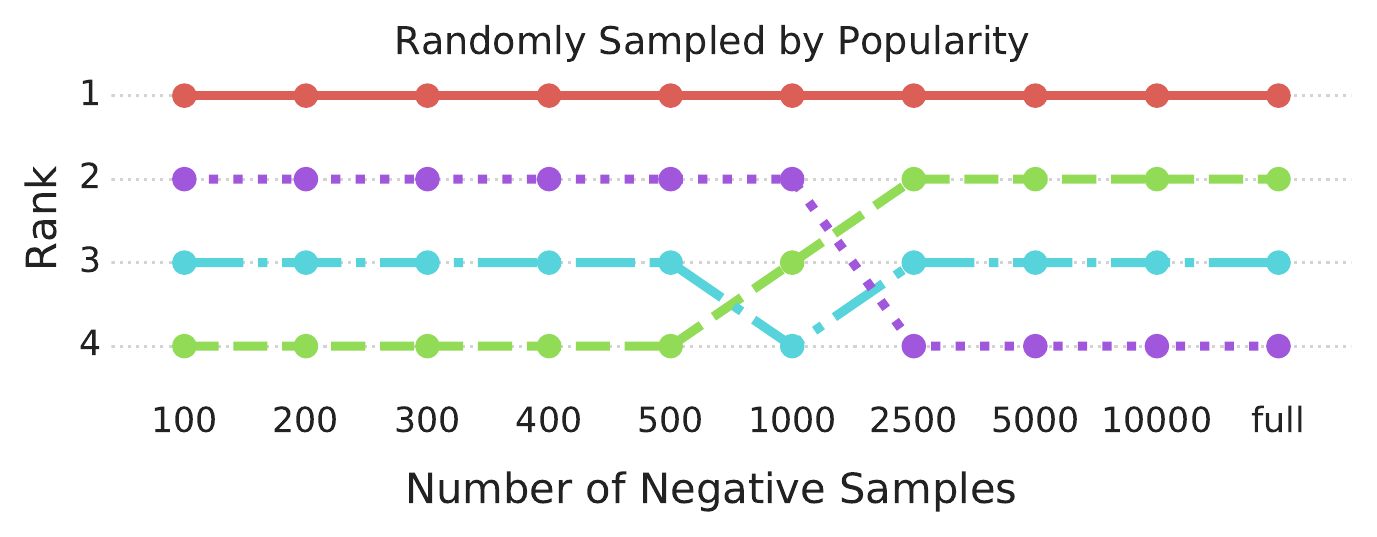}
  \caption{\steam}
  \label{fig:ranks-by-size-steam}
\end{subfigure}

\begin{subfigure}{\textwidth}
  \centering
  \includegraphics[width=0.4\linewidth]{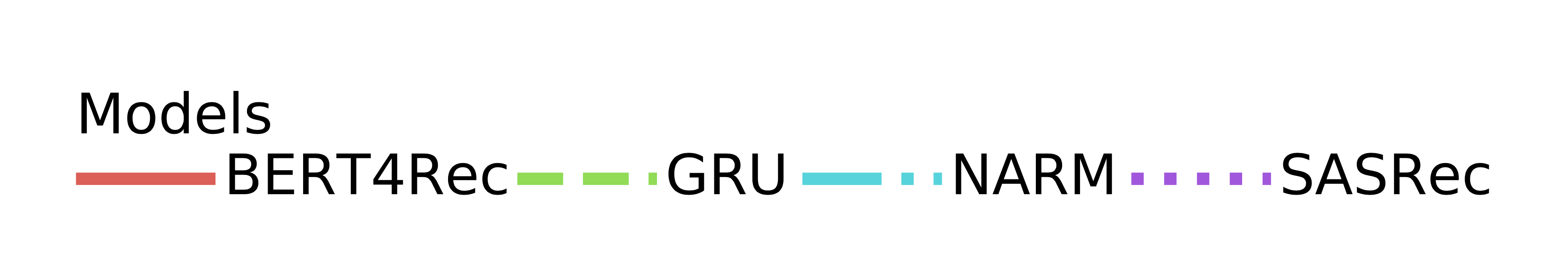}
  \end{subfigure}
  \Description[Changes in relative model ranking with varying sample size.]{Changes in the relative model ranking for \hrate{10} with increasing sample sizes for both uniform and popularity sampling.}
  \caption{Rankings of the four recommender models \gru, \narm, \sasrec and \bertforrec on the five considered datasets \ambeauty and Game, \movielenssystem \movielens and \movielenslarge and \steam for different negative sample sizes $\eta$.
  We vary the negative sample sizes from $100$ to "full", which means that we sample from the full item set, resulting in the \textit{\exact ranking}.
  On the left side, we plot the \textit{uniform} rankings and on the right side the \textit{popular} rankings.}
  \label{fig:ranks-by-size}
\end{figure*}

\subsection{Influence of Sample Size}

Previous work \citeauthorwitcite{Canmares20} showed that changing the \targetsize can have significant effects on the uniform model rankings.
The \exact ranking can be seen as the boundary case of any sampled evaluation where $\eta$ is equal to the number of items in the dataset.
In this case we can expect the approximation of the \exact ranking to improve with increasing \targetsize.
We calculate both sampled rankings for different \targetsize for all datasets and plot the rank changes in \Cref{fig:ranks-by-size} when using \hr{10} as the metric. 

Overall, we cannot observe many regularities from the plots.
The most striking observation is that increasing the sample size by one step only changes the rank by one place in most cases, hinting at a gradual change in recommendation performance with respect to sample size.
Additionally, the volatility of the ranking with respect to \targetsize seems to be highly dataset and, to some degree, sampling strategy dependent. 
For example, \Cref{fig:ranks-by-size-games} shows that for \amgame almost no changes in rank occur for both sampling rankings and for most sample sizes the \exact ranking is achieved.
However, the simple statistics of \amgame presented in \Cref{tab:dataset-stats} do not stand out and we cannot derive an explanation from the observed dataset properties.
In contrast, \ambeauty exhibits a high volatility with frequent ranking changes for both sampling rankings.
Even when almost half of the $\numprint{54542}$ items are sampled the ranking has not stabilized and still changes compared to the \exact ranking.
An instance where the behavior is different between sampling strategies is the \steam dataset in \Cref{fig:ranks-by-size-steam}. 
While uniform rankings are almost stable across different \targetsizes, popular rankings are more volatile, at least between $\eta = 500$ and $\eta = 2500$.
In summary, we cannot clearly identify a safe choice for $\eta$ for one of the rankings based on sampling nor for the dataset.

\section{Discussion And Limitations}
\label{sec:discussion}
In our study we trained four recently proposed neural sequential item recommendation models on five often used datasets and compared three rankings, which use different approaches to sample items for the target set used for calculating the metrics.
First, we wanted to investigate whether sampling by popularity produces model rankings that exhibit less inconsistencies regarding the \exact ranking and how it compares to uniform random sampling.
We found that both sampling strategies did not produce good approximations of the \exact ranking for most datasets, with the noteworthy exception of \amgame.
The obtained rankings on this dataset are for the most sample size consistent with each other.
Further analysis of why this is the case is necessary but that is out of the scope of this paper.
While sampling by popularity with $\eta = 100$ negative samples produced the model ranking established in recent work~\cite{sun2019bert4rec}, the \exact evaluation showed that \bertforrec does not hold state-of-the-art across all the datasets.
Our results suggest that \bertforrec profits from larger training corpora, since it outperforms all other models on the two larger datasets \movielenslarge and \steam using the \exact ranking, while it was unable to yield the best performance on the other, much smaller, datasets.
Further experiments (\eg hyperparameter studies to find maybe different optima for the metrics calculated on the full item set) must be carried out to analyse how the ranking of the model is really effected by the small datasets.
Nevertheless, these results call the current way of establishing state-of-the-art into question. 
Considering that the ranking depends on the selection for dataset, $\eta$ and how to sample the target set for metric calculation and that there is no consistent choice for $\eta$ that yields rankings consistent with the \exact ranking, the only safe choice for comparing models is to use the full item space as target set, although this may lead to longer runtimes for model evaluation.
Additionally, the results show that the clear picture, where one model outperforms all others across a range of datasets does not hold anymore if the \exact ranking is considered, at least for the considered datasets and models with their reported settings, as can be observed from the results for \bertforrec. 
At last, even the \exact ranking may not be the true ranking, since like all offline evaluation methods, it tends to underestimate the true metric score by assuming that unrated items are all non-relevant for the user~\cite{herlocker2004evaluating}.

Despite the extensive evaluation across models and datasets, our study has some limitations.
First, we are only comparing recent neural models based on Recurrent Neural Networks or Transformer architectures and do not consider Convolution Neural Networks based models like the one presented in~\cite{Tang18Caser} or other collaborative filtering methods or baselines. 
Therefore, it is possible that for some classes of models a different conclusion could be drawn.
Another limitation is the choice of datasets.
Although the review-based datasets have been popular in recent publications, there is a variety of datasets from a range of other domains available, that might exhibit different characteristics with impact on the sampled evaluation that were not considered in this evaluation.
Finally, we only used the common \textit{leave-one-out} dataset split that is often used with review datasets and did not consider alternative splits like a split by time.

\section{Conclusion}
\label{sec:conclusion}
In our study we focused on the impact of evaluation through sampling of small item sets on inconsistencies with the evaluation on the full item set when comparing recently published neural sequential item recommendation models.
While prior research~\cite{Canmares20} has studied inconsistencies between the ranking obtained using the full item set and the ranking obtained by uniform random sampling mostly on collaborative filtering models, our study extends this line of research in two key ways.
First, we include sampling by popularity, which is often used in recent work, as a second sampling strategy in our study and find that although it intuitively better approximates the item distribution in the dataset, it does not improve the consistency with the ranking obtained on the full item set.
Second, we study recent neural item recommendation models that can model complex interactions in the item sequences and the results of our experiments indicate that these models are also affected by the ranking inconsistencies when using sampling for the evaluation.
Our results strongly suggest that independent of the dataset, type of sampling or choice of the number of negative items $\eta$, a sampled evaluation will likely fail to approximate the ranking gained by considering the full item set correctly.
Therefore, it is a bad choice when comparing the performance of different sequential recommender models and cannot help to avoid calculating the metrics on the full item set. 

In future work we want to extend the study to more models to get better insights into the differences in performance utilizing the \exact ranking.
Additionally, we can extend the study using  different types of datasets.
We only used datasets constructed from user reviews or ratings, but other types of item sequence datasets exist (\eg user clicks in online stores), and might exhibit different characteristics.
Also we can study the effectiveness of already existing robust metrics introduced by~\citeauthorwitcite{Krichene20} on the evaluated models in this paper and different types of datasets.

\begin{acks}
This research was partly supported by the Bavarian State Ministry for Science and the Arts within the "Digitalisierungs\-zentrum für Präzisions- und Telemedizin" (DZ.PTM) project, as part of the master plan "BAYERN DIGITAL II".
\end{acks}

\bibliographystyle{ACM-Reference-Format}

\bibliography{bibliography}

\end{document}